\documentclass[acmsmall,screen,table]{acmart}

\AtBeginDocument{%
  }

\setcopyright{acmlicensed}
\acmJournal{PACMSE}
\acmYear{2025} \acmVolume{2} \acmNumber{ISSTA} \acmArticle{ISSTA049} \acmMonth{7}\acmDOI{10.1145/3728924}

\usepackage{numprint}
\usepackage{xspace}
\usepackage{amsfonts}
\usepackage{amsmath}
\usepackage{threeparttable}
\usepackage{listings}
\usepackage{enumitem}
\usepackage{subcaption}
\usepackage{tabularx}
\usepackage{multirow}
\usepackage{makecell}
\usepackage{pifont}
\usepackage{wrapfig}
\newcommand{\cmark}{\ding{51}}
\newcommand{\xmark}{\ding{55}}
\usepackage[linesnumbered,ruled,lined,vlined]{algorithm2e}
\SetArgSty{textnormal}

\newcommand*\wrapletters[1]{\wr@pletters#1\@nil}
\def\wr@pletters#1#2\@nil{#1\allowbreak\if&#2&\else\wr@pletters#2\@nil\fi}

\newcommand{\block}[1]{\href{https://etherscan.io/block/#1}{#1}\xspace}
\newcommand{\etherscanTx}[1]{\href{https://etherscan.io/tx/#1}{\wrapletters{#1}}\xspace}

\newcommand{\etherscanAddress}[1]{\href{https://etherscan.io/address/#1}{\wrapletters{#1}}\xspace}

\newcommand{\etal}{\textit{et al.}}
\npthousandsep{,}
\npdecimalsign{.}

\newcommand{\framework}{\textsc{Clue}\xspace}

\newcommand{\empirical}[1]{#1}

\newcommand{\datasetStartBlock}{\empirical{\block{9193268}}\xspace}
\newcommand{\datasetEndBlock}{\empirical{\block{14688629}}\xspace}
\newcommand{\datasetStartDate}{\empirical{1st of January, 2020}\xspace}
\newcommand{\datasetEndDate}{\empirical{30th of April, 2022}\xspace}
\newcommand{\datasetNotLabeledTxs}{\empirical{$\numprint{4786542}$}\xspace}

\newcommand{\totalLoC}{\empirical{$\numprint{6077}$}\xspace}
\newcommand{\traceSimulatorLoC}{\empirical{$\numprint{661}\ (11.7\%)$}\xspace}
\newcommand{\dyevmLoC}{\empirical{$\numprint{3019}\ (53.3\%)$}\xspace}
\newcommand{\graphLoC}{\empirical{$\numprint{1984}\ (35.0\%)$}\xspace}
\newcommand{\gremlinLoC}{\empirical{$\numprint{413}$}\xspace}

\newcommand{\reentrancyAttackTxs}{\empirical{$\numprint{87}$}\xspace}
\newcommand{\reentrancyGasTxs}{\empirical{$\numprint{1077}$}\xspace}
\newcommand{\reentrancyRandTxs}{\empirical{$\numprint{19985}$}\xspace}

\newcommand{\reentrancyNewTP}{\empirical{$\numprint{87}$}\xspace}
\newcommand{\reentrancyNewTPR}{\empirical{$\numprint{100}\%$}\xspace}
\newcommand{\reentrancyNewFN}{\empirical{$\numprint{0}$}\xspace}
\newcommand{\reentrancyNewFNR}{\empirical{$\numprint{0}\%$}\xspace}
\newcommand{\reentrancyNewGasTN}{\empirical{$\numprint{1069}$}\xspace}
\newcommand{\reentrancyNewGasTNR}{\empirical{$\numprint{99.26}\%$}\xspace}
\newcommand{\reentrancyNewGasFP}{\empirical{$\numprint{8}$}\xspace}
\newcommand{\reentrancyNewGasFPR}{\empirical{$\numprint{0.74}\%$}\xspace}
\newcommand{\reentrancyNewRandTN}{\empirical{$\numprint{19812}$}\xspace}
\newcommand{\reentrancyNewRandTNR}{\empirical{$\numprint{99.13}\%$}\xspace}
\newcommand{\reentrancyNewRandFP}{\empirical{$\numprint{173}$}\xspace}
\newcommand{\reentrancyNewRandFPR}{\empirical{$\numprint{0.87}\%$}\xspace}

\newcommand{\reentrancyReadNewFNR}{\empirical{$\numprint{0}\%$}\xspace}
\newcommand{\reentrancyReadNewGasTN}{\empirical{$\numprint{1073}$}\xspace}
\newcommand{\reentrancyReadNewGasTNR}{\empirical{$\numprint{99.63}\%$}\xspace}
\newcommand{\reentrancyReadNewGasFP}{\empirical{$\numprint{4}$}\xspace}
\newcommand{\reentrancyReadNewGasFPR}{\empirical{$\numprint{0.37}\%$}\xspace}
\newcommand{\reentrancyReadNewRandTN}{\empirical{$\numprint{19930}$}\xspace}
\newcommand{\reentrancyReadNewRandTNR}{\empirical{$\numprint{99.72}\%$}\xspace}
\newcommand{\reentrancyReadNewRandFP}{\empirical{$\numprint{55}$}\xspace}
\newcommand{\reentrancyReadNewRandFPR}{\empirical{$\numprint{0.28}\%$}\xspace}

\newcommand{\priceAttackTxs}{\empirical{$\numprint{53}$}\xspace}

\newcommand{\priceNewTP}{\empirical{$\numprint{51}$}\xspace}
\newcommand{\priceNewTPR}{\empirical{$\numprint{96.23}\%$}\xspace}
\newcommand{\priceNewFN}{\empirical{$\numprint{2}$}\xspace}
\newcommand{\priceNewFNR}{\empirical{$\numprint{3.77}\%$}\xspace}
\newcommand{\priceNewGasTN}{\empirical{$\numprint{1059}$}\xspace}
\newcommand{\priceNewGasTNR}{\empirical{$\numprint{98.51}\%$}\xspace}
\newcommand{\priceNewGasFP}{\empirical{$\numprint{16}$}\xspace}
\newcommand{\priceNewGasFPR}{\empirical{$\numprint{1.49}\%$}\xspace}
\newcommand{\priceNewRandTN}{\empirical{$\numprint{19886}$}\xspace}
\newcommand{\priceNewRandTNR}{\empirical{$\numprint{99.48}\%$}\xspace}
\newcommand{\priceNewRandFP}{\empirical{$\numprint{103}$}\xspace}
\newcommand{\priceNewRandFPR}{\empirical{$\numprint{0.52}\%$}\xspace}

\newcommand{\averageTraversalTime}{$169$ms\xspace}

\newcommand{\reentrancyAblationGasTNR}{$\numprint{82.24}\%$\xspace}
\newcommand{\reentrancyAblationGasFPR}{$\numprint{17.76}\%$\xspace}
\newcommand{\reentrancyAblationRandTNR}{$\numprint{99.03}\%$\xspace}
\newcommand{\reentrancyAblationRandFPR}{$\numprint{0.97}\%$\xspace}

\newcommand{\priceAblationAttackTPR}{$\numprint{100}\%$\xspace}
\newcommand{\priceAblationAttackFNR}{$\numprint{0}\%$\xspace}
\newcommand{\priceAblationGasTNR}{$\numprint{15.86}\%$\xspace}
\newcommand{\priceAblationGasFPR}{$\numprint{84.14}\%$\xspace}
\newcommand{\priceAblationRandTNR}{$\numprint{24.13}\%$\xspace}
\newcommand{\priceAblationRandFPR}{$\numprint{75.87}\%$\xspace}

\usepackage{acro}
\acsetup{single}

\DeclareAcronym{DeFi}{
  short = DeFi,
  long  = Decentralized Finance,
}
\newcommand{\DeFi}{\ac{DeFi}\xspace}

\DeclareAcronym{AMM}{
  short = AMM,
  long  = Automated Market Maker,
}

\DeclareAcronym{CPG}{
  short = CPG,
  long  = code property graph,
}
\newcommand{\CPG}{\ac{CPG}\xspace}

\DeclareAcronym{EPG}{
  short = EPG,
  long  = Execution Property Graph,
}
\newcommand{\EPG}{\ac{EPG}\xspace}

\DeclareAcronym{EVM}{
  short = EVM,
  long  = Ethereum virtual machine,
}
\newcommand{\EVM}{\ac{EVM}\xspace}

\DeclareAcronym{AST}{
  short = AST,
  long  = abstract syntax tree,
}
\newcommand{\AST}{\ac{AST}\xspace}

\DeclareAcronym{CFG}{
  short = CFG,
  long  = control-flow graph,
}
\newcommand{\CFG}{\ac{CFG}\xspace}

\DeclareAcronym{PDG}{
  short = PDG,
  long  = program dependence graph,
}
\newcommand{\PDG}{\ac{PDG}\xspace}

\DeclareAcronym{DCFG}{
  short = DCFG,
  long  = dynamic control-flow graph,
}
\newcommand{\DCFG}{\ac{DCFG}\xspace}

\DeclareAcronym{CTG}{
  short = CTG,
  long  = call trace graph,
}
\newcommand{\CTG}{\ac{CTG}\xspace}

\DeclareAcronym{VM}{
  short = VM,
  long  = virtual machine,
}
\newcommand{\VM}{\ac{VM}\xspace}
\newcommand{\VMs}{\acp{VM}\xspace}

\DeclareAcronym{EOA}{
  short = EOA,
  long  = externally owned account,
}
\newcommand{\EOA}{\ac{EOA}\xspace}

\DeclareAcronym{DApp}{
  short = DApp,
  long  = Decentralized Application,
}

\DeclareAcronym{IDS}{
  short = IDS,
  long  = intrusion detection system,
}

\DeclareAcronym{PtoP}{
  short = P2P,
  long  = peer-to-peer,
}
\newcommand{\PtoP}{\ac{PtoP}\xspace}

\DeclareAcronym{DDG}{
  short = DDG,
  long  = dynamic dependence graph,
}
\newcommand{\DDG}{\ac{DDG}\xspace}

\DeclareAcronym{PBS}{
  short = PBS,
  long  = Proposer/Builder Separation,
}

\DeclareAcronym{BSC}{
  short = BSC,
  long  = BNB Smart Chain,
}
\newcommand{\BSC}{\ac{BSC}\xspace}

\DeclareAcronym{LoC}{
  short = LoC,
  long  = lines of code,
}
\newcommand{\LoC}{\ac{LoC}\xspace}

\DeclareAcronym{TPR}{
  short = TPR,
  long  = true positive rate,
}
\newcommand{\TPR}{\ac{TPR}\xspace}

\DeclareAcronym{FPR}{
  short = FPR,
  long  = false positive rate,
}
\newcommand{\FPR}{\ac{FPR}\xspace}

\DeclareAcronym{TNR}{
  short = TNR,
  long  = true negative rate,
}
\newcommand{\TNR}{\ac{TNR}\xspace}

\DeclareAcronym{FNR}{
  short = FNR,
  long  = false negative rate,
}
\newcommand{\FNR}{\ac{FNR}\xspace}

\DeclareAcronym{SOTA}{
  short = SOTA,
  long = state-of-the-art
}
\newcommand{\SOTA}{\ac{SOTA}\xspace}

\newcommand{\Toutl}[1][l]{\textsc{Out}_{#1}}
\newcommand{\TR}[1][\mathcal{T}]{\textsc{R}_{#1}}

\definecolor{codegreen}{rgb}{0,0.6,0}
\definecolor{codegray}{rgb}{0.5,0.5,0.5}
\definecolor{codepurple}{rgb}{0.58,0,0.82}
\definecolor{backcolour}{rgb}{0.95,0.95,0.92}

\lstdefinelanguage{Solidity}{
	keywords=[1]{anonymous, assembly, assert, balance, break, call, callcode, case, catch, class, constant, continue, constructor, contract, debugger, default, delegatecall, delete, do, else, emit, event, experimental, export, external, false, finally, for, function, gas, if, implements, import, in, indexed, instanceof, interface, internal, is, length, library, log0, log1, log2, log3, log4, memory, modifier, new, payable, pragma, private, protected, public, pure, push, require, return, returns, revert, selfdestruct, send, solidity, storage, struct, suicide, super, switch, then, this, throw, transfer, true, try, typeof, using, value, view, while, with, addmod, ecrecover, keccak256, mulmod, ripemd160, sha256, sha3, unchecked}, 
	keywordstyle=[1]\color{magenta}\bfseries,
	keywords=[2]{address, bool, byte, bytes, bytes1, bytes2, bytes3, bytes4, bytes5, bytes6, bytes7, bytes8, bytes9, bytes10, bytes11, bytes12, bytes13, bytes14, bytes15, bytes16, bytes17, bytes18, bytes19, bytes20, bytes21, bytes22, bytes23, bytes24, bytes25, bytes26, bytes27, bytes28, bytes29, bytes30, bytes31, bytes32, enum, int, int8, int16, int24, int32, int40, int48, int56, int64, int72, int80, int88, int96, int104, int112, int120, int128, int136, int144, int152, int160, int168, int176, int184, int192, int200, int208, int216, int224, int232, int240, int248, int256, mapping, string, uint, uint8, uint16, uint24, uint32, uint40, uint48, uint56, uint64, uint72, uint80, uint88, uint96, uint104, uint112, uint120, uint128, uint136, uint144, uint152, uint160, uint168, uint176, uint184, uint192, uint200, uint208, uint216, uint224, uint232, uint240, uint248, uint256, var, void, ether, finney, szabo, wei, days, hours, minutes, seconds, weeks, years},	
	keywordstyle=[2]\color{codepurple}\bfseries,
	keywords=[3]{block, blockhash, coinbase, difficulty, gaslimit, number, timestamp, msg, data, gas, sender, sig, value, now, tx, gasprice, origin},	
	keywordstyle=[3]\color{teal}\bfseries,
	identifierstyle=\color{black},
	sensitive=false,
	comment=[l]{//},
	morecomment=[s]{/*}{*/},
	commentstyle=\color{gray}\ttfamily,
	stringstyle=\color{red}\ttfamily,
	morestring=[b]',
	morestring=[b]"
}

\lstdefinelanguage{Decompiled}{
	keywords=[1]{anonymous, assembly, assert, balance, break, call, callcode, case, catch, class, constant, continue, constructor, contract, debugger, default, delegatecall, delete, do, else, emit, event, experimental, export, external, false, finally, for, function, gas, if, implements, import, in, indexed, instanceof, interface, internal, is, length, library, log0, log1, log2, log3, log4, memory, modifier, new, payable, pragma, private, protected, public, pure, push, require, return, returns, revert, selfdestruct, send, solidity, storage, struct, suicide, super, switch, then, this, throw, transfer, true, try, typeof, using, value, view, while, with, addmod, ecrecover, keccak256, mulmod, ripemd160, sha256, sha3}, 
	keywordstyle=[1]\color{magenta}\bfseries,
	keywords=[2]{address, bool, byte, bytes, bytes1, bytes2, bytes3, bytes4, bytes5, bytes6, bytes7, bytes8, bytes9, bytes10, bytes11, bytes12, bytes13, bytes14, bytes15, bytes16, bytes17, bytes18, bytes19, bytes20, bytes21, bytes22, bytes23, bytes24, bytes25, bytes26, bytes27, bytes28, bytes29, bytes30, bytes31, bytes32, enum, int, int8, int16, int24, int32, int40, int48, int56, int64, int72, int80, int88, int96, int104, int112, int120, int128, int136, int144, int152, int160, int168, int176, int184, int192, int200, int208, int216, int224, int232, int240, int248, int256, mapping, string, uint, uint8, uint16, uint24, uint32, uint40, uint48, uint56, uint64, uint72, uint80, uint88, uint96, uint104, uint112, uint120, uint128, uint136, uint144, uint152, uint160, uint168, uint176, uint184, uint192, uint200, uint208, uint216, uint224, uint232, uint240, uint248, uint256, var, void, ether, finney, szabo, wei, days, hours, minutes, seconds, weeks, years},	
	keywordstyle=[2]\color{teal}\bfseries,
	keywords=[3]{block, blockhash, coinbase, difficulty, gaslimit, number, timestamp, msg, data, gas, sender, sig, value, now, tx, gasprice, origin},	
	keywordstyle=[3]\color{codepurple}\bfseries,
	identifierstyle=\color{black},
	sensitive=false,
	comment=[l]{//},
	morecomment=[s]{/*}{*/},
	commentstyle=\color{codegreen}\ttfamily,
	stringstyle=\color{red}\ttfamily,
	morestring=[b]',
	morestring=[b]"
}

\lstdefinelanguage{Opcode}{
	keywords=[1]{REVERT,JUMPDEST,SELFBALANCE,CALLER,PUSH20,EQ,DUP1,PUSH1,PUSH2,PUSH4,JUMPI,POP,MSTORE,ADDRESS,GAS,STATICCALL,CALL},
	keywordstyle=[1]\color{teal}\bfseries,
	keywords=[2]{PC, Disassembled, Code},
	keywordstyle=[2]\color{codegray}\ttfamily,
	comment=[l]{//},
	morecomment=[s]{/*}{*/},
	commentstyle=\color{gray}\ttfamily,
}

\definecolor{delim}{RGB}{20,105,176}
\definecolor{numb}{RGB}{106, 109, 32}
\definecolor{string}{rgb}{0.64,0.08,0.08}

\lstdefinelanguage{json}{
    numbers=left,
    numberstyle=\footnotesize,
    frame=single,
    rulecolor=\color{black},
    showspaces=false,
    showtabs=false,
    breaklines=true,
    postbreak=\raisebox{0ex}[0ex][0ex]{\ensuremath{\color{gray}\hookrightarrow\space}},
    breakatwhitespace=true,
    basicstyle=\ttfamily\footnotesize,
    upquote=true,
    morestring=[b]",
    stringstyle=\color{string},
    literate=
     *{0}{{{\color{numb}0}}}{1}
      {1}{{{\color{numb}1}}}{1}
      {2}{{{\color{numb}2}}}{1}
      {3}{{{\color{numb}3}}}{1}
      {4}{{{\color{numb}4}}}{1}
      {5}{{{\color{numb}5}}}{1}
      {6}{{{\color{numb}6}}}{1}
      {7}{{{\color{numb}7}}}{1}
      {8}{{{\color{numb}8}}}{1}
      {9}{{{\color{numb}9}}}{1}
      {\{}{{{\color{delim}{\{}}}}{1}
      {\}}{{{\color{delim}{\}}}}}{1}
      {[}{{{\color{delim}{[}}}}{1}
      {]}{{{\color{delim}{]}}}}{1},
}

\newif\ifcomments
\commentstrue
\ifcomments
\newcommand{\todo}[1]{\textcolor{orange}{TODO: #1}}

\newcommand{\dawn}[1]{\textcolor{red}{dawn: #1}}
\newcommand{\zhe}[1]{\textcolor{teal}{zhe: #1}}
\newcommand{\kaihua}[1]{\textcolor{blue}{zhe: #1}}
\else
\newcommand{\todo}[1]{}

\newcommand{\dawn}[1]{}
\newcommand{\zhe}[1]{}
\newcommand{\kaihua}[1]{}
\fi

\newcommand{\revision}[1]{#1}

\begin{document}

\title{Enhancing Smart Contract Security Analysis with Execution Property Graphs}

%

\author{Kaihua Qin}
\authornote{Both authors contributed equally to this research.}
\authornote{Also affiliated with Berkeley Center for Responsible, Decentralized Intelligence (RDI)}
\authornote{Also affiliated with Decentralized Intelligence AG}
\affiliation{%
  \institution{Yale University}
  \country{USA}
}
\email{kaihua@qin.ac}

\author{Zhe Ye}
\authornotemark[1] 
\authornotemark[2]
\affiliation{
  \institution{UC Berkeley}
  \country{USA}
}
\email{zhey@berkeley.edu}

\author{Zhun Wang}
\authornotemark[2]
\affiliation{
  \institution{UC Berkeley}
  \country{USA}
}
\email{zhun.wang@berkeley.edu}

\author{Weilin Li}
\affiliation{
  \institution{University College London}
  \country{UK}
}
\email{weilin.li.24@ucl.ac.uk}

\author{Liyi Zhou}
\authornotemark[2]
\authornotemark[3]
\affiliation{
  \institution{The University of Sydney}
  \country{Australia}
}
\email{liyi.zhou@sydney.edu.au}

\author{Chao Zhang}
\affiliation{
  \institution{Tsinghua University}
  \country{China}
}
\email{chaoz@tsinghua.edu.cn}

\author{Dawn Song}
\authornotemark[2]
\affiliation{
  \institution{UC Berkeley}
  \country{USA}
}
\email{dawnsong@cs.berkeley.edu}

\author{Arthur Gervais}
\authornotemark[2]
\authornotemark[3]
\affiliation{
  \institution{University College London}
  \country{UK}
}
\email{a.gervais@ucl.ac.uk}
%

\begin{abstract}
Smart contract vulnerabilities have led to significant financial losses, with their increasing complexity rendering outright prevention of hacks increasingly challenging. This trend highlights the crucial need for advanced forensic analysis and real-time intrusion detection, where dynamic analysis plays a key role in dissecting smart contract executions. Therefore, there is a pressing need for a unified and generic representation of smart contract executions, complemented by an efficient methodology that enables the modeling and identification of a broad spectrum of emerging attacks.

We introduce \framework, a dynamic analysis framework specifically designed for the Ethereum virtual machine. Central to \framework is its ability to capture critical runtime information during contract executions, employing a novel graph-based representation, the Execution Property Graph. A key feature of \framework is its innovative graph traversal technique, which is adept at detecting complex attacks, including (read-only) reentrancy and price manipulation. Evaluation results reveal \framework's superior performance with high true positive rates and low false positive rates, outperforming state-of-the-art tools. Furthermore, \framework's efficiency positions it as a valuable tool for both forensic analysis and real-time intrusion detection.
\end{abstract}




\maketitle

\acresetall

\section{Introduction}
Despite considerable efforts devoted to securing smart contracts, the community continues to experience attacks that lead to annual cumulative losses surpassing billions of US dollars~\cite{zhou2022sok}.
Contract audits are a prevalent practice in the blockchain and the \DeFi community to prevent exploits prior to contract deployments. These audits typically combine automated tools, such as static analysis~\cite{tsankov2018securify,feist2019slither}, symbolic execution~\cite{mossberg2019manticore,zheng2022park}, and fuzzing~\cite{jiang2018contractfuzzer,nguyen2020sfuzz}, with manual assessments by security auditors. However, while these pre-deployment tools are effective in mitigating some types of bugs, they often overlook the runtime state of smart contracts and fail to capture the unpredictable interactions between different contracts, thereby missing complex vulnerabilities. Moreover, the manual assessment process is inherently prone to human error, further exacerbating the challenge of ensuring comprehensive smart contract security.

\begin{figure*}[t]
    \centering
    \includegraphics[width=0.86\textwidth]{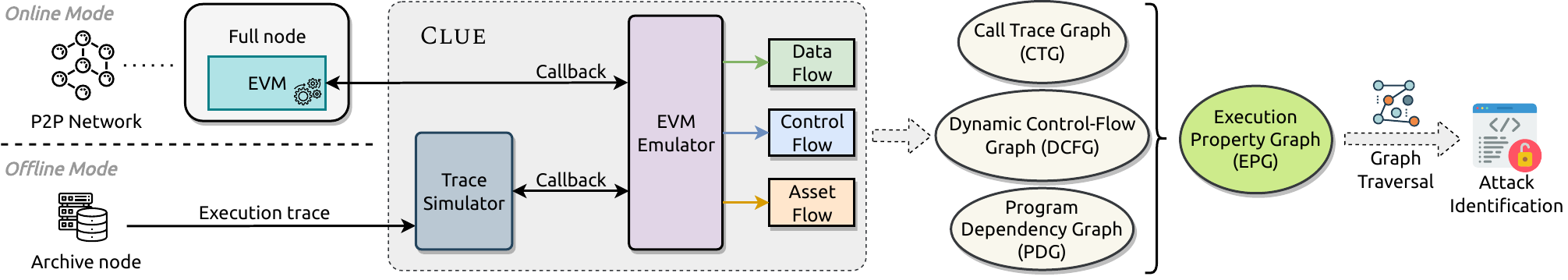}
    \caption{
    \revision{High-level architecture of \framework.
\framework supports both real-time (online) detection of unconfirmed transactions and postmortem (offline) analysis of historical traces. At its core, an \acs{EVM} emulator instruments the runtime to capture detailed execution data, control flow, and asset flow, which are then combined into our unified \acs{EPG} (cf.\ Section~\ref{sec:contract-execution-representation}). A graph traversal engine (cf.\ Section~\ref{sec:traversal}) then identifies potential attacks from the constructed \acs{EPG}. Notably, the transaction trace simulator replays each execution trace so that it appears indistinguishable from a real-time \acs{EVM} run, allowing the emulator interface to remain consistent across both online and offline modes (cf.\ Section~\ref{sec:impl} for details).}
    }
    \label{fig:architecture}
\end{figure*}

The complete elimination of vulnerabilities prior to smart contract deployment remains a significant challenge. This reality underscores the critical role of post-attack forensic analysis in smart contract security. Forensic analysis is essential for understanding attack mechanics and identifying specific vulnerabilities, a crucial step given that blockchain user assets may remain vulnerable and still be susceptible to exploitation even after an initial attack~\cite{qin2021attacking}. Moreover, similar vulnerabilities might exist in other smart contracts, posing an ongoing risk~\cite{zhou2022sok}. Therefore, a comprehensive and efficient forensic analysis is imperative to minimize the chances of further losses. Such analysis necessitates an in-depth examination of specific blockchain transactions, including contract executions and their interactions, where the application of dynamic analysis techniques is particularly advantageous. Not limited to postmortem scenarios, dynamic analysis also serves as a proactive security measure. It can be used to scrutinize pending transactions before their confirmation, effectively functioning as an advanced intrusion detection system.

Existing dynamic analysis solutions for smart contract security are often tailored to specific attacks~\cite{grossman2017online,rodler2019sereum}. This specialization, while valuable for detecting specific attack patterns, restricts their flexibility and extensibility in addressing a broad range of issues. In contrast, some other solutions that aim for a more generic approach fall short in precise data-flow tracking~\cite{chen2020soda}, thereby omitting critical runtime information for fine-grained analysis. In this context, the need for a unified representation of contract executions becomes paramount. A fine-grained, comprehensive model that encapsulates the complexities of smart contract interactions and behaviors in a coherent structure is crucial. Such a representation not only facilitates a deeper understanding of contract dynamics but also serves as a foundation for advanced security analyses.

\revision{
To bridge these gaps, we introduce \framework (cf.\ Figure~\ref{fig:architecture}), a generic dynamic analysis framework for \acf{EVM} smart contracts. \framework streamlines the security analysis process; it operates at the granularity of individual transactions by tracking contract executions and collecting runtime data~---~such as dynamic data, control flow, and asset flow~---~throughout each transaction. Specifically, \framework accepts either live transactions or historical traces as its input, which it tracks in an instrumented environment to capture essential execution information for security analysis. Leveraging these insights, \framework constructs our novel \acf{EPG}~---~a unified representation that interweaves the call hierarchy, asset transfers, control-flow structure, and data dependencies of every transaction. Once the \EPG is built, \framework applies a graph traversal mechanism to efficiently detect potential smart contract attacks encoded in specific traversal rules. Through this approach, \framework functions both as a high-speed forensic analysis tool and a real-time intrusion detection system, providing swift identification of suspicious behaviors. Our methodology, centered on the \EPG representation, enables concise descriptions and precise identification of contract attacks, thereby elevating the effectiveness of dynamic analysis in smart contract security.
}

Although various graph-based techniques are well-established in program analysis, their use in dynamic analysis specifically for smart contract security remains relatively underexplored. Prior studies have developed specialized graphs tailored to detect specific types of attacks; for instance, Rodler \etal~\cite{rodler2019sereum} and Wu \etal~\cite{wu2023defiranger} each design distinct, customized graph structures aimed at detecting different categories of attacks. However, to the best of our knowledge, we are the first to propose a generic graph representation, the \EPG, capable of identifying multiple attack types. While certain attack types may still necessitate supplementary information, as demonstrated in Section~\ref{sec:evaluation}, the extensibility of \framework allows for easy integration of such data without requiring structural modifications. This unified representation provides a foundation for ongoing analysis, facilitating rapid adaptation to emerging attack patterns without the need to reinvent graph structures for each new attack. This responsiveness is crucial for the proactive defense and for advancing smart contract security analysis.

We outline the key contributions of our work as follows.

\begin{itemize}
    \item We develop a dynamic analysis framework, \framework, for the \EVM smart contracts. \framework is designed to track both intra- and inter-contract executions, gathering runtime information that includes data, control, and asset flows.

    \item We introduce the \EPG~---~a comprehensive and unified graph-based representation that effectively captures the behaviors of smart contract executions, facilitating efficient contract security analysis.

    \item We introduce an innovative graph traversal technique based on \EPG to identify potential smart contract attacks. This approach extracts semantic information from the low-level \EPG representation of transactions, automating the detection of malicious patterns without requiring access to the original source code. In this work, we specifically design and implement two graph traversal instances targeting reentrancy attacks (including the recently identified read-only variant) and price manipulation attacks.

    \item Our traversal-based security analysis achieves a \TPR of~\reentrancyNewTPR and~\priceNewTPR for reentrancy and price manipulation, respectively. Meanwhile, it maintains a \FPR of~\reentrancyNewRandFPR and~\priceNewRandFPR for these attacks. In comparison to \SOTA dynamic analysis tools for reentrancy and price manipulation, \framework stands out in terms of both efficiency and accuracy. Our evaluations underscore the efficiency of \framework, which on average completes the analysis of a transaction in~\averageTraversalTime. This performance highlights \framework's potential as a real-time intrusion detection system. We further show that \framework successfully detects the read-only reentrancy attack, a recently disclosed attack, which eludes detection by the related work.

\end{itemize}

\section{Background}

\subsection{Blockchain and Decentralized Finance}
The advent of Bitcoin~\cite{bitcoin} marked the beginning of a new era of decentralized databases known as blockchains. A blockchain is a \PtoP distributed database that consists of a series of blocks, each containing a list of transactions. In a blockchain, accounts are represented through unique addresses, and an address can claim ownership of data attributed to it. Every transaction represents a change in the state of the blockchain, such as a transfer of ownership or funds.
In addition to simple transfers of funds, transactions can carry quasi-Turing-complete transition functions, which are realized in the form of so-called smart contracts. Smart contracts in essence are programs running on top of blockchains and allow the implementation of various applications, including financial products, i.e., \DeFi. These \DeFi applications enable a plethora of use cases, such as financial exchanges~\cite{xu2023sok}, lending~\cite{qin2021empirical}, and flash loans~\cite{qin2021attacking}.

\subsection{Ethereum Virtual Machine}
Smart contracts are typically executed within a \VM environment. One of the most popular \VMs for this purpose is the \EVM, which we focus on within this work. \EVM supports two types of accounts: \textit{(i)} the \EOA, which is controlled by a cryptographic private key and can be used to initiate transactions, and \textit{(ii)} the smart contract, which is bound to immutable code and can be invoked by other accounts or contracts.

The code for an \EVM contract is usually written in a high-level programming language (e.g., Solidity), and then compiled into \EVM bytecode. When users want to invoke the execution of a smart contract, they can send a transaction from their \EOA to the target contract address, including any necessary parameters or data. An invoked contract can also call other smart contracts within the same transaction, enabling complex interactions between different contracts. \EVM utilizes three main components to execute a smart contract: the stack, memory, and storage. The stack and memory are volatile areas, which are reset with each contract invocation, for storing and manipulating data, while the storage is persistent across multiple executions.

While \EVM was originally created for the Ethereum blockchain, it has been adopted by a variety of blockchains beyond Ethereum, including \BSC and Avalanche. Throughout this work, we focus on the context of Ethereum, but it is worth noting that the contract execution representation \EPG and the dynamic analysis framework \framework can also be applied to other EVM-compatible blockchains.

\subsection{Smart Contract Security}
Smart contracts can have bugs and vulnerabilities, just like programs written for traditional systems. Both the academic and the industry communities have therefore adopted tremendous efforts at securing smart contracts. Most of the efforts to date have focused on the smart contract layer, where manual and automated audits aim to identify bugs before a smart contract's deployment. Despite these efforts, smart contract vulnerabilities have resulted in billions of dollars in losses~\cite{zhou2022sok}. As we will explore in this work, detecting and investigating attacks, forensics in general, is a tedious manual effort that we aim to simplify.

\section{Execution Property Graph}\label{sec:contract-execution-representation}
\begin{figure*}[t]
\centering
    \resizebox{0.9\textwidth}{!}{
        \begin{subfigure}[b]{0.30\textwidth}
             \centering
             \includegraphics[width=\textwidth]{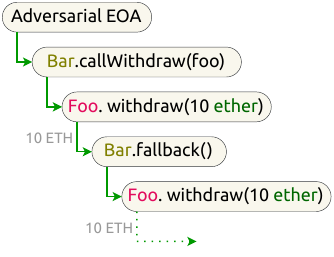}
             \caption{Call Trace Graph}
             \label{fig:call-trace-example}
         \end{subfigure}
         \hfill
         \begin{subfigure}[b]{0.30\textwidth}
             \centering
             \includegraphics[width=\textwidth]{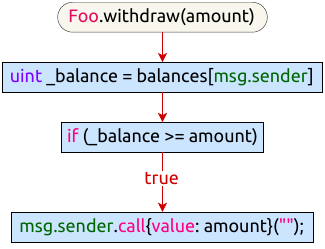}
             \caption{Dynamic Control Flow Graph}
             \label{fig:dcfg-example}
         \end{subfigure}
         \hfill
         \begin{subfigure}[b]{0.40\textwidth}
             \centering
             \includegraphics[width=\textwidth]{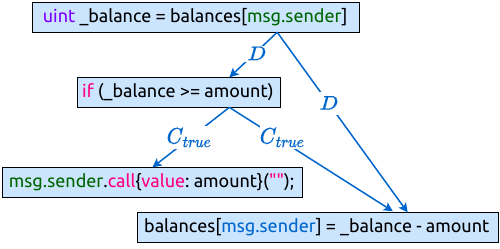}
             \caption{Dynamic Dependence Graph}
             \label{fig:ddg-example}
         \end{subfigure}
    }
    \caption{Contract execution representations for the reentrancy example in Listing~\ref{lst:reentrancy}. In Figure~\ref{fig:ddg-example}, $C$ and $D$ indicate control and data dependency respectively.}
    \label{fig:contract-execution-representations}
\end{figure*}

Code representation is a well-studied topic in program analysis literature~\cite{binkley2007source}. Classic code representations, including \AST~\cite{aho2007compilers}, \CFG~\cite{allen1970control}, and \PDG~\cite{ferrante1987program}, are also applicable to smart contract analysis. Despite their effectiveness in identifying particular contract vulnerabilities~\cite{ren2021making,luu2016making,tsankov2018securify}, these static representations ignore the dynamic information exposed in the concrete contract executions, which we aim to capture in this work. Specifically, in the following, we demonstrate (cf.\ Figure~\ref{fig:contract-execution-representations}) how to represent smart contract executions with the \CTG, \DCFG, and \DDG. We then illustrate the construction of the \EPG by consolidating these three fundamental representations.

\definecolor{raspberry}{rgb}{0.89, 0.04, 0.36}
\newcommand{\FooContract}{\texttt{\color{raspberry}Foo}\xspace}
\newcommand{\BarContract}{\texttt{\color{olive}Bar}\xspace}

\subsection{Property Graph}
A property graph is a multi-relational graph, where vertices and edges are attributed with a set of
\begin{wrapfigure}{r}{0.54\columnwidth}
\begin{lstlisting}[belowskip=0pt,label=lst:reentrancy,language=Solidity, caption={Reentrancy vulnerability example. The contract \FooContract contains a reentrancy vulnerability (line~\ref{line:withdrawsendether} and~\ref{line:storageupdate}), which an attacker can exploit with the  contract \BarContract.},escapechar=@,basicstyle=\scriptsize\ttfamily]
pragma solidity ^0.8.0;
contract Foo {
  mapping (address => uint) public balances;
  function withdraw(uint amt) public {
    uint _balance = balances[msg.sender];
    if (_balance >= amt) {@\label{line:ifcondition}@
      msg.sender.call{value: amt}("");@\label{line:withdrawsendether}@
      balances[msg.sender] = _balance - amt;@\label{line:storageupdate}@
    } else {
      revert("insufficient balance");@\label{line:revert}@
    }
  }
}
contract Bar {
  function callWithdraw(address foo) public {
    Foo(foo).withdraw(10 ether);
  }
  fallback() external payable {
    if (address(this).balance < 99999 ether) {
      callWithdraw(msg.sender);
    }
  }
}
\end{lstlisting}
\vspace{-30pt}
\end{wrapfigure}
key-value pairs, known as properties~\cite{rodriguez2012graph}.
Contrary to a single-relational graph, where edges are homogeneous in meaning, edges in a
property graph are labeled and thus heterogeneous. Properties grant a graph the ability to represent non-graphical data (e.g., different types of relationships between entities in a social network graph). A property graph is formally defined in Definition~\ref{def:property-graph}.
\begin{definition}[Property Graph]\label{def:property-graph}
A property graph is defined as $G = (V, E, \lambda, \mu)$, where $V$ is a set of vertices and $E\subseteq (V\times V)$ is a set of directed edges. $\lambda: E\rightarrow\Sigma$ is an edge labeling function that labels edges with symbols from the alphabet $\Sigma$, while $\mu: (V \cup E)\times K\rightarrow S$ assigns key-value properties to vertices and edges, where $K$ is a set of property keys and $S$ is a set of property values.
\end{definition}

In this work, $K^V$ and $K^E$ denote the property key sets of vertices and edges respectively, s.t.\ $K=K^V \cup K^E$. We use $S^{\textit{k}}$ to denote the set of property values associated with the property key $\textit{k}\in K$.

\subsection{Running Example}\label{sec:running-example}
As a running example, we consider a thoroughly studied contract vulnerability, \emph{reentrancy}, which results in the infamous ``The DAO'' attack~\cite{cecchetti2021compositional,bose2022sailfish}. Listing~\ref{lst:reentrancy} features two contracts, the vulnerable contract \FooContract and the adversarial contract \BarContract.
\FooContract allows users to keep a balance of ETH (the native
cryptocurrency on Ethereum) at their address, and also allows withdrawing the deposited ETH. The \texttt{withdraw} function, however, performs the withdraw transaction prior to deducting the account balance (line~\ref{line:withdrawsendether} and~\ref{line:storageupdate}, Listing~\ref{lst:reentrancy})~---~and is hence vulnerable to reentrancy.
\BarContract exploits the reentrancy through repeated, reentrant calls. An adversary calls the \texttt{callWithdraw} function of \BarContract and \BarContract further calls the \texttt{withdraw} function of \FooContract. In the \texttt{withdraw} function, when \FooContract sends the specified amount of ETH to \BarContract (cf.\ line~\ref{line:withdrawsendether}, Listing~\ref{lst:reentrancy}), the \texttt{fallback} function of \BarContract is triggered. This \texttt{fallback} function invocation allows repeated ETH withdrawals from \BarContract.

Note that the graphs in Figure~\ref{fig:contract-execution-representations} may appear trivial to generate by parsing the transaction execution trace with the smart contract source code. However, in practice, contracts are not always open-source, especially malicious ones such as \BarContract. Consequently, it is more practical and robust to build smart contract execution representations based on the bytecode instead of any high-level language. Although Figure~\ref{fig:contract-execution-representations} presents Solidity for clarity in examples, we clarify that our contract execution representation \EPG bases on the \EVM bytecode and does not require access to the contract source code.

\subsection{Call Trace Graph}\label{sec:call-trace-representation}
In a transaction, multiple smart contracts can be invoked in a nested and successive manner. We show in our reentrancy example that the contract \FooContract and \BarContract invoke each other repeatedly. Such contract invocations can be structured into a \CTG (cf.\ Figure~\ref{fig:call-trace-example}). Given a transaction, the \CTG captures the sequence and hierarchy of contract invocations. Ignoring the detailed executions within a contract, every invocation in a \CTG is abstracted into a quintuple: \textit{(i)} from address~---~caller; \textit{(ii)} to address~---~callee; \textit{(iii)} invocation opcode~---~the triggering opcode (\texttt{CALL}, \texttt{DELEGATECALL}, \texttt{CALLCODE}, \texttt{STATICCALL}); \textit{(iv)} call value~---~the amount of ETH transferred with the call; \textit{(v)} call data~---~call parameters.


Owing to its simplicity, the concept of \CTG is extensively employed in contract security analysis, particularly for manual attack postmortem processes. Popular \EVM transaction decoders, such as \href{https://ethtx.info/}{ethtx.info}, essentially display the \CTG of a given transaction to enhance interpretation.

Formally, a \CTG property graph is defined by $G_T = (V_T, E_T, \lambda_T, \mu_T)$. In a \CTG, vertices $V_T$ correspond to contracts, while edges $E_T$ represent invocations from a caller vertex to a callee vertex (cf.\ Figure~\ref{fig:epg}). It is important to note that $\mu_T$ assigns the \textit{asset flow} property to $e_T\in E_T$ when $e_T$ is associated with asset transfers. There are generally two types of assets in \EVM: \textit{(i)} the native cryptocurrency ETH, and \textit{(ii)} assets realized by smart contracts (e.g., fungible tokens). The property value of \textit{asset flow} includes the transferred \textit{asset}, \textit{from} and \textit{to} address, as well as the transfer \textit{amount}. \revision{In Figure~\ref{fig:epg}, we present an asset flow property of~$10$ ether from a victim contract \texttt{0xf0..f0} to an attack contract \texttt{0xba..ba} during a reentrant call, which can be used to indicate a reentrancy attack.}

\subsection{Dynamic Control-Flow Graph}\label{sec:dcfg-representation}
A \CFG represents smart contract code as a graph, where each vertex denotes a basic block. A basic block is a piece of contract code that is executed sequentially without a jump. Basic blocks are connected with directed edges, representing code jumps in the control flow. 

A \CFG is a static representation of contract code, but can also be constructed dynamically while a contract executes, i.e., a so-called \DCFG. Figure~\ref{fig:dcfg-example} shows the \DCFG of the \texttt{Foo} contract (cf.\ Listing~\ref{lst:reentrancy}) during the reentrancy exploit. Contrary to the \CFG, a \DCFG focuses on the dynamic execution information, hence ignoring the unvisited basic blocks and jumps. For instance, the code at line~\ref{line:revert}, Listing~\ref{lst:reentrancy} is not included in the \DCFG (cf.\ Figure~\ref{fig:dcfg-example}).
Formally, a \DCFG can be defined as $G_C = (V_T \cup V_C, E_C, \lambda_C, \mu_C)$, where $V_T$ and $V_C$ represent the contract and basic block vertices respectively. Edges $E_C$ indicate code jumps between basic blocks (cf.\ Figure~\ref{fig:epg}).

\subsection{Dynamic Dependence Graph}\label{sec:ddg-representation}
The \PDG is another form of code representation outlining the dependency relationship in a program. There are two main types of dependencies, data flow dependency and control flow dependency~\cite{horwitz1992use}. An instruction $X$ has a flow dependency on an instruction $Y$, if $Y$ defines a value used by $X$. Note that data flow dependencies are transitive, i.e., if $Y$ is dependent on $Z$ and $X$ is dependent on $Y$, then $X$ is dependent on $Z$.
For control dependency, informally, an instruction $X$ has a control dependency on a branching instruction $Y$, if changing the branch target for $Y$ may cause $X$ not to be executed.

Similar to how a \DCFG is constructed from concrete executions, a \PDG can also be built dynamically, which is referred to as a \DDG. We present the \DDG of our reentrancy example in Figure~\ref{fig:ddg-example}, where $D$ and $C$ denote data and control dependency, respectively.

\begin{figure}[t]
    \centering
    \includegraphics[width=0.65\columnwidth]{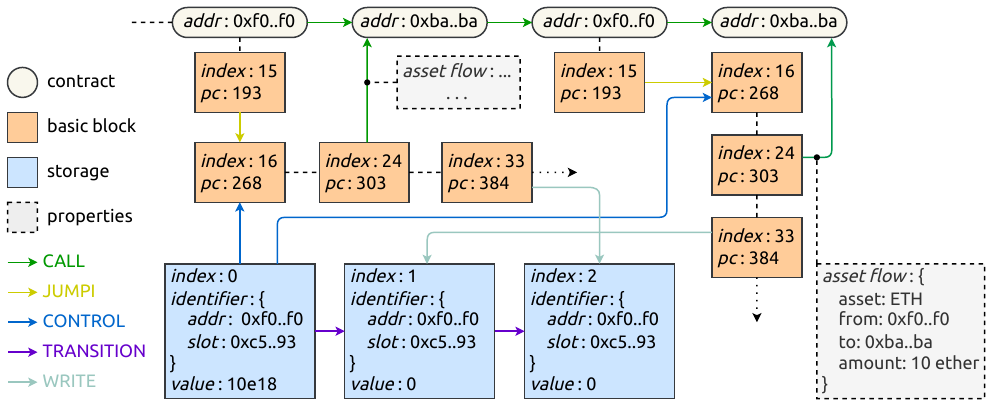}
    \caption{\revision{Partial \EPG of the reentrancy attack transaction (cf.\ Section~\ref{sec:running-example}). The \EPG is a composite structure formed by merging three fundamental property graphs: \CTG, \DCFG, and \DDG. The \CTG comprises contract vertices interconnected by \texttt{CALL} edges, which may be assigned with \textit{asset flow} properties. A \DCFG is initialized with a contract vertex, followed by basic block vertices linked via \texttt{JUMPI} edges. Building upon the \DCFG, the \DDG introduces data source vertices (e.g., storage) and incorporates edges (e.g., \texttt{CONTROL}) to represent data and control dependencies. 
    }}
    \label{fig:epg}
\end{figure}

The \DDG is built upon the \DCFG. Formally, a \DDG $G_D = (V_T\cup V_C\cup V_D, E_D, \lambda_D, \mu_D)$ contains contract vertices $V_T$, basic block vertices $V_C$, and data source (e.g., storage) vertices $V_D$, while edges $E_D$ represent the data and control dependencies (cf.\ Figure~\ref{fig:epg}).

\revision{
As shown in Figure~\ref{fig:epg}, a \texttt{TRANSITION} edge connects two data source vertices whenever the value of a data source is updated. A \texttt{WRITE} edge links the basic block vertex $v_T$, which executes the data writing operation (e.g., \texttt{SSTORE}), to the updated data source vertex $v_D$. A control dependency edge~---~labeled \texttt{CONTROL}~---~connects the data source vertex of a \texttt{JUMPI} condition to the corresponding target basic block vertex. Furthermore, if the content written by one data source depends on another, they are connected by a \texttt{DEPENDENCY} edge. Although \texttt{DEPENDENCY} edges are not shown in Figure~\ref{fig:epg}, they are widely used in the price manipulation traversal (cf.\ Figure~\ref{fig:price_traversal}, Section~\ref{sec:price-oracle}).
}



\subsection{Constructing the Execution Property Graph}\label{sec:construct-epg}
The \EPG is constructed by merging the three basic property graphs (cf.\ Definition~\ref{def:epg}). Because every \DCFG originates from a contract vertex and the \DDG is built upon the \DCFG, the graph merging process is straightforward.

\begin{definition}[Execution Property Graph]\label{def:epg}
An \EPG $G=(V, E, \lambda, \mu)$ is constructed by merging \CTG, \DCFG, and \DDG, s.t., 
\begin{equation}\label{eq:epg_model}
    \begin{aligned}
        & V = V_T \cup V_C \cup V_D \qquad E = E_T \cup E_C \cup E_D \cup \mathbf{T}(E_T)\\
        & \lambda = \lambda_T \cup \lambda_C \cup \lambda_D\qquad \mu = \mu_T \cup \mu_C \cup \mu_D \qquad \Sigma = \Sigma_T \cup \Sigma_C \cup \Sigma_D\\
    \end{aligned}
\end{equation}
\revision{where \(\lambda = \lambda_T \cup \lambda_C \cup \lambda_D\) means that for each edge \(e\in E\), \(\lambda(e)=\lambda_T(e)\) if \(e\in E_T\), \(\lambda(e)=\lambda_C(e)\) if \(e\in E_C\), and \(\lambda(e)=\lambda_D(e)\) if \(e\in E_D\) (and similarly for \(\mu = \mu_T \cup \mu_C \cup \mu_D\)). $\mathbf{T}: V_T\times V_T \rightarrow V_C\times V_T$ is a transformation function, elaborated further in the following.}
\end{definition}

To incorporate more execution details into the \EPG, we apply the transformation function $\mathbf{T}$ to the \CTG edges $E_T$. For every $e_T\in E_T$, $\mathbf{T}$ generates a new edge $e'_T$ and inserts it into the \EPG. 
The label and properties of $e'_T$ are inherited from $e_T$, while the tail vertex of $e'_T$ is changed from the caller contract vertex to the basic block vertex initiating the contract invocation. The generated edges enable the \EPG to capture contract invocations in a more granular manner.

Figure~\ref{fig:epg} presents the \EPG of the reentrancy attack transaction (cf. Section~\ref{sec:running-example}), with unnecessary details omitted. From the diagram, it is evident that the balance update (i.e., the storage write) occurs after the ETH transfer, constituting the root cause of this reentrancy vulnerability.

\section{Traversals-based Security Analysis}\label{sec:traversal}
The \EPG provides extensive information about the contract executions involved in a transaction. In this section, we explore how graph traversals, a prevalent method for mining information in property graphs, identify contract attacks. 

\subsection{Rationale}

The literature suggests that the primary objective of an individual executing a smart contract attack is typically to obtain financial gain~\cite{zhou2022sok}. By exploiting vulnerabilities, an attacker may illicitly acquire financial assets in a way that deviates from the intended design of the compromised contract. As a result, an attack transaction often entails the transfer of assets from the victim to the attacker. To perform a comprehensive security analysis, it is crucial to identify suspicious asset transfers within a transaction. More specifically, a profound understanding of the root cause of a contract attack necessitates an in-depth examination of the underlying mechanisms by which the attacker procures assets from the victim.

For example, in the context of a reentrancy attack (cf. Section~\ref{sec:running-example}), the attacker can repeatedly invoke a vulnerable contract in a reentrant fashion. To detect such an attack, it is vital to pinpoint the specific asset transfers associated with the reentrant contract calls. Moreover, a reentrancy attack exploits the inconsistent contract state, which is modified subsequent to the malicious asset transfer. The extensive runtime information contained in the \EPG enables the detection of suspicious execution patterns, such as reentrant contract invocations and inconsistent contract states. This identification can be achieved by traversing the graph and conducting an appropriate search, as further elaborated in Section~\ref{sec:reentrancy}.

We therefore propose \framework{}, a framework employs \EPG traversals to enable automated transaction security analysis.
The goal of traversing the \EPG is to infer high-level semantic information from low-level graphic representations and subsequently identify malicious logic patterns. This inference process does not necessitate the knowledge of the application-level logic, rendering the methodology more generic and extensible. Nonetheless, for specific attacks, such as price manipulation, relying solely on graph traversal may result in high \FPR and \FNR. To address these attacks, our methodology also supports integrating corresponding domain knowledge (e.g., estimating price change to detect price manipulation attack) into the traversal process, which can substantially decrease the \FPR and \FNR. We present the details in Section~\ref{sec:evaluation}.

\subsection{Traversal Details}
\revision{
In the following, we delve into the traversal specifics for the two most prevalent real-world attacks: reentrancy and price manipulation.
To ease explanation, we define the transitive closure function as outlined in Equation~\ref{eq:transitiveClosure}, which returns all vertices in $V$ from which some vertex in $S$ \emph{can reach} by following a sequence of edges labeled with any label from a set $\Lambda$ representing specific edge types.
\begin{equation}\label{eq:transitiveClosure}
\begin{aligned}
\texttt{transitiveClosure}(S,\, \Lambda,\, V) =\ \Bigl\{\, v \in V \;\Bigm|\; \exists\, s \in S,\; \exists\, k \ge 1,\; \exists\, v_1,\dots,v_k \in V :\\ v_0 = s,\quad v_k = v, 
\quad \forall\, i \in \{0,\dots,k-1\},\; \lambda((v_i, v_{i+1})) \in \Lambda \Bigr\}
\end{aligned}
\end{equation}
Similarly, $\texttt{revClosure}(S,\, \Lambda,\, V)$ represents all vertices in $V$ from which some vertex in $S$ \emph{is reachable} by following a sequence of edges labeled with any label from $\Lambda$.
}



\subsubsection{Reentrancy}\label{sec:reentrancy}
The concept of reentrancy vulnerability has been well explored in prior research. Specifically, the reentrancy attack is defined by three key characteristics: \textit{(i)} the presence of reentrant contract calls, \textit{(ii)} the control of asset transfers relying on outdated storage values, and \textit{(iii)} storage updates subsequent to asset transfers~\cite{rodler2019sereum}.

\begin{figure*}[t]
\centering
    \resizebox{\textwidth}{!}{
        \begin{subfigure}[b]{0.35\textwidth}
             \centering
             \includegraphics[width=\textwidth]{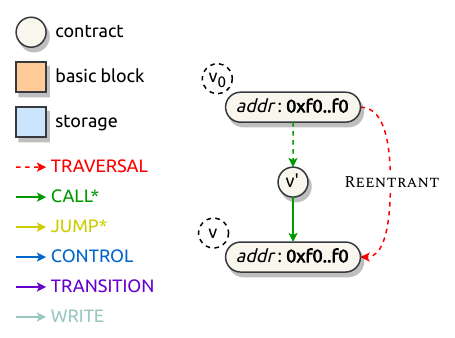}
             \caption{\textsc{Reentrant} Rule}
             \label{fig:reentrancy_traversal_1}
         \end{subfigure}
         \hfill
         \begin{subfigure}[b]{0.35\textwidth}
             \centering
             \includegraphics[width=\textwidth]{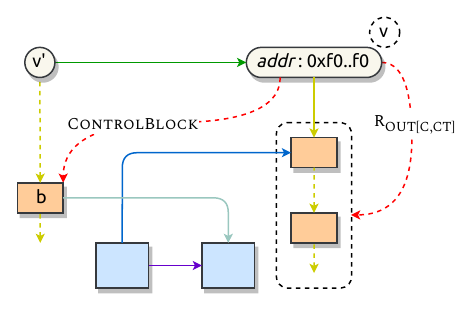}
             \caption{\textsc{ControlBlock} Rule}
             \label{fig:reentrancy_traversal_2}
         \end{subfigure}
         \hfill
         \begin{subfigure}[b]{0.25\textwidth}
             \centering
             \includegraphics[width=\textwidth]{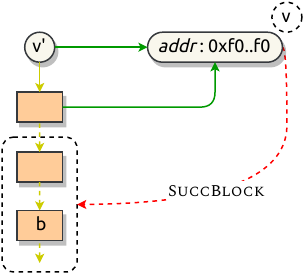}
             \caption{\textsc{SuccBlock} Rule}
             \label{fig:reentrancy_traversal_3}
         \end{subfigure}
    }
    \caption{Three traversal rules of the reentrancy attack detection. \texttt{CALL*} represents all edge labels of $E_T$ and $\mathbf{T}(E_T)$, \texttt{JUMP*} represents all edges of $E_C$. Except \texttt{TRAVERSAL}, dotted edges represent one or more edges with internal vertices omitted. \texttt{TRAVERSAL} edges represent the result of traversal execution. $\TR[{\Toutl[C, CT]}]$ denotes a recursive traversal following outgoing edges of $E_C$ and $\mathbf{T}(E_T)$. 
    }
    \label{fig:reentrancy_traversal}
\end{figure*}

\revision{
We therefore propose a method for identifying reentrancy attacks based on three traversal rules (i.e., \textsc{Reentrant}, \textsc{ControlBlock}, and \textsc{SuccBlock}) that capture the aforementioned characteristics. These rules are illustrated in Figure~\ref{fig:reentrancy_traversal} and formalized in Algorithm~\ref{alg:reentrancy}. Specifically, \textsc{Reentrant} navigates the \texttt{CALL} edges, identifying pairs of contract vertices with identical contract addresses to detect reentrant calls (cf.\ line~\ref{line:reentrant}, Algorithm~\ref{alg:reentrancy} and Figure~\ref{fig:reentrancy_traversal_1}). For each identified reentrant pair, \textsc{ControlBlock} (cf.\ line~\ref{line:controlblock_start}--\ref{line:controlblock_end}, Algorithm~\ref{alg:reentrancy} and Figure~\ref{fig:reentrancy_traversal_2}) traverses the \DCFG to locate the basic block vertices of the reentrant call that are associated with asset flows. Additionally, \textsc{ControlBlock} searches for storage vertices updated within the parent calls (via \texttt{WRITE} edges) while controlling the basic blocks in the reentrant call (via \texttt{CONTROL} edges), which indicate that the asset transfer potentially relies on an outdated value. Note that storage vertices connected by \texttt{TRANSITION} edges represent the same storage variable being updated. Moreover, \textsc{SuccBlock} ensures that the storage update occurs after the reentrant call (cf.\ line~\ref{line:succblock_start}--\ref{line:succblock_end}, Algorithm~\ref{alg:reentrancy} and Figure~\ref{fig:reentrancy_traversal_3}). We remark that, while Figure~\ref{fig:reentrancy_traversal} presents the three rules separately for clarity, they are in practice executed simultaneously rather than in isolation.
}

\begin{algorithm}[t]
    \small
    \DontPrintSemicolon
    \caption{\revision{Reentrancy Traversal}}\label{alg:reentrancy}
    \SetKwInOut{Input}{Input}
    \SetKwInOut{Output}{Output}

    \Input{EPG $G = (V, E, \lambda, \mu)$}
    \Output{True if a reentrancy attack is detected; False otherwise}
    
    \ForEach{$(u, v) \in V_T \times V_T $ such that 
        $\mu(u, \textit{addr}) = \mu(v, \textit{addr})$ \textbf{and} 
        $\exists\, k \geq 2,\ \exists\, v_0, v_1, \ldots, v_k \in V_T$ with 
        $v_0 = u$, $v_k = v$, and $\forall\, i \in \{0, 1, \dots, k-1\},\ \lambda((v_i,v_{i+1})) = \texttt{CALL}$}{\label{line:reentrant}
            $\mathcal{B} \gets \text{The set of basic block vertices of } v \text{ that are associated with asset flows}$\;\label{line:controlblock_start}
            \(\mathcal{D} \gets \{\, d \in V_D \mid \exists\, b \in \mathcal{B}:\ \lambda((d, b)) = \texttt{CONTROL} \,\}\)\;
            \(\mathcal{D}_{\texttt{old}} \gets \texttt{revClosure}(\mathcal{D},\, \{\texttt{TRANSITION}\},\, V_D)\)\;
            \(\mathcal{W} \gets \{\, w \mid \exists\, d \in \mathcal{D}_{\texttt{old}}:\ \lambda((w, d)) = \texttt{WRITE} \,\}\)\;\label{line:controlblock_end}
            \ForEach{$w \in \mathcal{W}$}{\label{line:succblock_start}
                \lIf{$w \text{ executes after the invocation of } v $}{\Return{true}}\label{line:succblock_end}
            }
        }
    \Return{false}\;
\end{algorithm}


Notably, our traversal approach comprehensively addresses all three types of classic reentrancy attacks (i.e., cross-contract, delegated, create-based reentrancy) as discussed in~\cite{rodler2019sereum}, owing to the \EPG's unified model that captures all forms of contract calling.

\subsubsection{Price Manipulation}\label{sec:price-oracle} Price manipulation attacks in smart contracts typically involve exploiting vulnerabilities to manipulate asset prices, often by tampering with price oracles. In cases where a price manipulation attack occurs within a single transaction, the attacker can alter the price, typically stored in the contract storage, and subsequently profit by, for instance, exchanging assets at the manipulated price. This type of attack entails an asset transfer, the amount of which is influenced by the manipulated storage.

\begin{figure}[t]
    \centering
    \begin{minipage}{0.55\columnwidth}
        \centering
        \includegraphics[width=\linewidth]{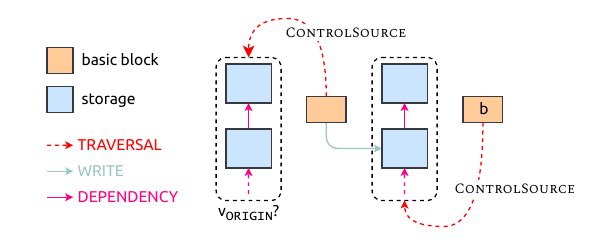}
    \end{minipage}%
    \hfill
    \begin{minipage}{0.45\columnwidth}
        \captionof{figure}{Overview of \textsc{WriteControl} traversal used in price manipulation detection. \textsc{WriteControl} identifies asset flows (denoted by the basic block \texttt{b} in the figure) that depend on some data sources manipulated without authentication. $v_{\texttt{ORIGIN}}?$ checks if the sender address is the data source of the storage variables.}
        \label{fig:price_traversal}
    \end{minipage}
\end{figure}

\revision{Figure~\ref{fig:price_traversal} visualizes our price manipulation traversal, termed \texttt{WriteControl}, which is outlined in Algorithm~\ref{alg:price}.
On a high level, the traversal identifies asset transfers that are ``influenced'' by prior storage changes and check whether these changes are authenticated. The process involves two stages of the \textsc{ControlSource} traversal that identify all control data sources associated with a given basic block. Let \(V_{\texttt{transfer}}\) denote the set of all basic block vertices associated with asset transfers. In the first stage (cf. lines~\ref{line:price_control_source_vt_start}--\ref{line:price_control_source_vt_end}, Algorithm~\ref{alg:price}), \textsc{ControlSource} is applied to \(V_{\texttt{transfer}}\) to collect every data source governing asset transfers throughout the transaction. Next, we reverse-traverse the \texttt{WRITE} edge (cf. line~\ref{line:price_rev_traverse_write}, Algorithm~\ref{alg:price}) to pinpoint the basic block vertices that manipulate these data sources. We verify whether the manipulation is authenticated by checking if the transaction sender’s data source, \(v_{\texttt{ORIGIN}}\), is among the collected control sources. This is done by applying \textsc{ControlSource} again on the basic block vertices identified as manipulating asset transfers (cf. lines~\ref{line:price_control_source_man_start}--\ref{line:price_control_source_man_end_and_check}, Algorithm~\ref{alg:price}). If \(v_{\texttt{ORIGIN}}\) is absent, we conclude that unauthenticated price manipulation has likely occurred.
}


\begin{algorithm}[t]
    \small
    \DontPrintSemicolon
    \caption{\revision{Price Manipulation Traversal}}\label{alg:price}
    \SetKwInOut{Input}{Input}
    \SetKwInOut{Output}{Output}

    \Input{EPG $G = (V, E, \lambda, \mu)$}
    \Output{Price manipulation attack detection result}
    
    \(\mathcal{B} \gets \texttt{transitiveClosure}(V_{\texttt{transfer}},\, \{\texttt{CALL}, \texttt{JUMP}\},\, V_T \cup V_C)\)\; \label{line:price_control_source_vt_start}
    \(\mathcal{D} \gets \{\, d \mid \exists\, b \in \mathcal{B} :\  \lambda((d, b)) = \texttt{CONTROL} \,\}\)\;
    \(\mathcal{D}^{+} \gets \texttt{revClosure}(\mathcal{D},\, \{\texttt{DEPENDENCY}\},\, V_D)\)\; \label{line:price_control_source_vt_end}
    \(\mathcal{B}_{\texttt{manipulate}} \gets \{\, b \mid \exists\, d \in \mathcal{D}^{+} :\  \lambda((b, d)) = \texttt{WRITE} \,\}\)\; \label{line:price_rev_traverse_write}
    \(\mathcal{B}_{\texttt{manipulate}}^{+} \gets \texttt{transitiveClosure}(\mathcal{B}_{\texttt{manipulate}},\, \{\texttt{CALL}, \texttt{JUMP}\},\, V_T \cup V_C)\)\; \label{line:price_control_source_man_start}
    \(\mathcal{D}_{\texttt{manipulate}} \gets \{\, d \mid \exists\, b \in \mathcal{B}_{\texttt{manipulate}}^{+} :\  \lambda((d, b)) = \texttt{CONTROL} \,\}\)\;
    \(\mathcal{D}_{\texttt{manipulate}}^{+} \gets \texttt{revClosure}(\mathcal{D}_{\texttt{manipulate}},\, \{\texttt{DEPENDENCY}\},\, V_D)\)\;
    \Return{\(v_{\texttt{ORIGIN}} \notin \mathcal{D}_{\texttt{manipulate}}^{+}\)} \label{line:price_control_source_man_end_and_check}
\end{algorithm}

\section{Experimental Evaluation}\label{sec:evaluation}


\revision{
In this section, we evaluate \framework to answer the following research questions.
\begin{description}
    \item[RQ1] How effectively does \framework detect the targeted vulnerabilities~---~reentrancy and price manipulation~---~in terms of true/false positives and negatives?
    \item[RQ2] Is \framework efficient enough for real-time deployment, considering its runtime overhead?
    \item[RQ3] How does \framework's detection capability compare to \SOTA dynamic analysis tools?
    \item[RQ4] To what extent does the integrated \EPG structure enhance detection accuracy compared to partial representations (omitting \CTG, \DCFG, or \DDG)?
\end{description}
}

\revision{
For RQ1, we measure \framework's detection accuracy on real-world transactions by analyzing standard metrics including \FPR and \FNR. For RQ2, we assess \framework's performance overhead by evaluating its average detection time per transaction, thereby determining its viability for real-time deployment. For RQ3, we compare \framework with \SOTA detection tools, focusing on accuracy and runtime overhead, on the same datasets to ensure a fair comparison.  Finally, for RQ4, we conduct an ablation study by removing single constituent graph from the integrated \EPG and observing the impact on the accuracy and detection time, highlighting the necessity of the unified representation.
}

\subsection{Evaluation Setup and Datasets}

\subsubsection{Implementation and Setup}\label{sec:impl}
\revision{We implement a prototype of the \framework framework (cf.\ Figure~\ref{fig:architecture}), which includes the \EVM emulator, \EPG construction, and graph traversal. Notably, the emulator instruments the \EVM to capture precise runtime information~---~encompassing dynamic data, control flow, and asset flow~---~crucial for constructing the \EPG. It leverages go-ethereum's built-in tracer interface to provide an \EVM callback at runtime, thereby minimizing code changes to the \EVM and removing the need to transmit large (often gigabyte-scale) JSON logs for \EPG extraction. This design enhances efficiency in online detection. Meanwhile, the trace simulator parses the transaction execution trace and triggers the callback function as though the transaction were running on a live \EVM. As a result, the emulator's interface can be reused seamlessly for offline scenarios, where standard transaction logs can be downloaded from any blockchain node without requiring node-side modifications. This makes integration into existing infrastructures straightforward.
}

\revision{
Our \framework prototype is implemented in Golang with a total of \totalLoC~\LoC. Specifically, the trace simulator comprises \traceSimulatorLoC~\LoC, and the EVM emulator adds \dyevmLoC~\LoC. The graph construction module contributes \graphLoC~\LoC. We employ \href{https://github.com/apache/tinkerpop/}{Apache TinkerGraph} as the graph backend, connected to \framework via a local WebSocket interface. The graph traversal module uses the Golang binding of the Gremlin~\cite{rodriguez2015gremlin} query language, which consists of \gremlinLoC~\LoC.

The evaluation runs on an Ubuntu v$22.04$ machine with~$24$ CPU cores and $128$ GB of RAM.
}


\begin{table*}[t]
\centering
\caption{Evaluation datasets.}
\label{tab:evaluation-dataset}
\scalebox{0.8}{%
\begin{threeparttable}
\begin{tabularx}{\textwidth}{ccX}
\toprule
Category & Dataset & Description\\ \midrule
\multirow{2}{*}{\textbf{Attack dataset}} & \multirow{2}{*}{\emph{Attack}} & The \emph{Attack} dataset comprises all attack transactions of the evaluated attack type from the dataset provided by~\cite{zhou2022sok}. \\
\midrule
\multirow{12.25}{*}{\makecell{\textbf{Non-attack}\\\textbf{datasets}}} &\multirow{6}{*}{\emph{High-Gas}} & As attack transactions often consume more gas due to complex actions, the \emph{High-Gas} dataset comprises the top~$\numprint{1100}$ transactions with the highest gas\tnote{1} consumption that interact with victim contracts, excluding the attack transactions. \emph{High-Gas} represents complex execution logic, evaluating performance on benign transactions that could be misclassified as attacks. \\
\cmidrule{2-3}
& \multirow{5}{*}{\emph{Regular}} & The \emph{Regular} dataset contains $\numprint{20000}$ randomly selected transactions that interact with victim contracts, with attacks emitted. This dataset is representative of typical non-attack transactions while maintaining the potential for false positives due to their interaction with vulnerable contracts.\\
\bottomrule
\end{tabularx}
\begin{tablenotes}
\item[1] \small{unit of computation cost on Ethereum (cf. \url{https://ethereum.org/en/developers/docs/gas/})}
\end{tablenotes}
\end{threeparttable}
}
\end{table*}

\subsubsection{Evaluation Dataset}~\label{sec:evaluation-dataset}
To evaluate the effectiveness and performance of \framework, we construct a dataset derived from a reference dataset presented in~\cite{zhou2022sok}, which comprises~$\numprint{2452}$ attack transactions on Ethereum. Our evaluation focuses on the two most prevalent types of attacks, \textit{(i)} reentrancy and \textit{(ii)} price manipulation. These attacks are represented by~\reentrancyAttackTxs and~\priceAttackTxs attack transactions, respectively, in our dataset.

We further enrich the evaluation dataset by collecting all transactions that have engaged with the victim contracts in the aforementioned attack transactions. The collection spans from block \datasetStartBlock (\datasetStartDate) to \datasetEndBlock (\datasetEndDate), resulting in~\datasetNotLabeledTxs transactions that are not explicitly labeled as attacks and serve for comparative analysis. Specifically, for each attack type, we create three sub-datasets, \emph{Attack}, \emph{High-Gas}, and \emph{Regular} (cf.\ Table~\ref{tab:evaluation-dataset}).

\subsubsection{Comparison Dataset}\label{sec:benchmark-dataset}
In our evaluation, we emphasize the importance of focusing on real-world transactions to ensure the practical applicability of our findings. 
Detecting reentrancy and price manipulation attacks with dynamic analysis has been extensively studied in the literature~\cite{rodler2019sereum,zhang2020txspector,wu2023defiranger,chen2024flashsyn,kong2023defitainter,xie2024defort}, providing a basis for comparing \framework with \SOTA approaches. However, for the reentrancy evaluation, due to the lack of open-source implementations or incompatibility of available open-source implementations with the latest version of \EVM, we are unable to apply these existing solutions to our evaluation dataset. Instead, we trace all~$\numprint{77987922}$ transactions in the first~$\numprint{4500000}$ blocks of Ethereum to create a comparison dataset, which is also used for evaluation in~\textsc{Sereum}~\cite{rodler2019sereum} and as a benchmark in~\textsc{TxSpector}~\cite{zhang2020txspector}.
For price manipulation evaluation, we also utilize the $D1$ comparison dataset from \textsc{DeFort}~\cite{xie2024defort}, which comprises 54 confirmed price manipulation attacks across multiple chains including Ethereum, BSC, Polygon, and Fantom.
This dataset has been used to evaluate several \SOTA price manipulation detection tools~\cite{xie2024defort}.

\subsection{RQ1: Detection Accuracy}
\subsubsection{Reentrancy}
Table~\ref{tab:reentrancy-eval} presents the evaluation results for identifying reentrancy attacks on the \textit{Attack}, \textit{High-Gas}, and \textit{Regular} datasets. \framework demonstrates efficacy in detecting reentrancy attacks, achieving a zero \FNR in the \textit{Attack} dataset, along with low \FPR of~$0.74\%$ and~$0.87\%$ in \textit{High-Gas} and \textit{Regular}, respectively.
Notably, \framework successfully discovers~$11$ \textit{new} reentrancy transactions within the \textit{Regular} dataset. These transactions are confirmed as true positives that had been missed in the \textit{Attack} dataset from~\cite{zhou2022sok}. We have reported these missed true positives to the authors, who have acknowledged the findings and confirmed their intention to address them in the next revision of their work.

\paragraph{Reentrancy Refinements}
In practice, we have observed attackers executing storage-change-based reentrancy attacks, aimed at manipulating the internal storage of victim contracts instead of transferring assets.
Such attacks, which do not involve asset flow in the reentrant call, are not detected by the traversal rule as initially defined in Section~\ref{sec:reentrancy}.
However, the flexible design of \framework allows for the refinement of traversal rules to accurately recognize and respond to these evolving attack patterns.
In our evaluation, we patch the rule to include the detection of unexpected storage modifications in the reentrant calls, significantly broadening the detection scope of the original traversal rule.

\begin{table}[t]
\centering
\caption{Reentrancy evaluation. The \textit{Attack} dataset showcases a high true positive rate (\reentrancyNewTPR) and a low false negative rate (\reentrancyNewFNR). In non-attack datasets, the true negative rates are remarkably high (\reentrancyNewGasTNR) for \textit{High-Gas} and (\reentrancyNewRandTNR) for \textit{Regular}. The few false positives are caused by Flash Loan and Rebase Token cases.}\label{tab:reentrancy-eval}
\rowcolors{2}{gray!10}{white}
\scalebox{0.78}{%
\begin{tabular}{l|c|cc}
\toprule
            &  & \multicolumn{2}{c}{Non-attack} \\ 
Dataset     & \textit{Attack} & \textit{High-Gas} & \textit{Regular} \\ 
\midrule
Size            & \reentrancyAttackTxs & \reentrancyGasTxs  & \reentrancyRandTxs \\ 
Gas Cost       & 3.33 $\pm$ 3.41M & 2.13 $\pm$ 1.38M & 0.24 $\pm$ 0.29M \\ 
Traversal Time & 0.32 $\pm$ 0.93s  & 52 $\pm$ 109ms & 16 $\pm$ 347ms \\ 
TP (\%)             & \reentrancyNewTP (\reentrancyNewTPR)  & - & -\\
FN (\%)             & \reentrancyNewFN (\reentrancyNewFNR)    & - & - \\ 
TN (\%)             & -       & \reentrancyNewGasTN (\reentrancyNewGasTNR) & \reentrancyNewRandTN (\reentrancyNewRandTNR)\\
FP (\%)             & -       & \reentrancyNewGasFP (\reentrancyNewGasFPR) & \reentrancyNewRandFP (\reentrancyNewRandFPR)\\
\bottomrule
\end{tabular}
}
\end{table}

\paragraph{Unreported Vulnerability Discovery -- imBTC Reentrancy}\label{sec:imBTC-case-study}
After further investigation of these $11$ newly discovered reentrancy transactions, we discover a potential vulnerability in token \emph{imBTC}.\footnote{imBTC: \etherscanAddress{0x3212b29E33587A00FB1C83346f5dBFA69A458923}} The attacks under investigation commence with the utilization of imBTC to exchange for ETH within the Uniswap V1 pool. Subsequently, the attacker exploits the callback function during the transfer of imBTC as an ERC777 token, initiating a reentrancy attack. This enables the attacker to execute another exchange with inaccurate pricing prior to the liquidity pool update, ultimately yielding a profit. The fundamental cause of these attacks can be ascribed to the discrepancy between the Uniswap V1 standard and the ERC777 standard. Through further analysis of imBTC, our research uncovers potential attack vectors within Uniswap V2 as well. It is crucial to clarify that this vulnerability does not stem directly from the Uniswap smart contracts themselves but rather from the incompatibilities introduced by the functionalities of ERC777 tokens. In accordance with responsible disclosure practices, we have reported these findings to the developers and have received prompt responses.\footnote{As the vulnerability has not yet been addressed, we are unable to provide further details in this submission.}

\paragraph{False Positive Analysis for Non-Attack Dataset}\label{sec:fp-analysis}
In analyzing the \textit{High Gas} and \textit{Regular} datasets, \framework identifies~\reentrancyNewGasFP out of~\reentrancyGasTxs and~\reentrancyNewRandFP out of~\reentrancyRandTxs benign transactions as reentrancy attacks, respectively (i.e., false positives). Consequently, the \FPR amounts to~\reentrancyNewGasFPR and~\reentrancyNewRandFPR in the \textit{High-Gas} and \textit{Regular} datasets, respectively. Both false-positive cases from the \textit{High-Gas} dataset are associated with the flash loan process in Euler Finance.\footnote{\url{https://www.euler.finance/}} The callback function tied to the flash loan creates a reentrant call pattern. Furthermore, there are special variables related to the state of borrowing and repayment. They are changed both when borrowing and repayment, and affect the control flow and the asset flow. This issue can be resolved by verifying the \texttt{from} and \texttt{to} addresses of the suspicious asset flow.
The single false-positive case from the \textit{Regular} dataset also interacts with another rebase token similar to imBTC as we described in Section~\ref{sec:imBTC-case-study}. It triggered a reentrant call from Uniswap V2 router when performing a transfer to swap and add liquidity. However, there is no potential vulnerability in this situation as the caller is limited to Uniswap V2 router.

\subsubsection{Price Manipulation}\label{sec:price-manipulation-rq1}

\begin{table}[t]
\centering
\caption{Price manipulation evaluation. \framework achieves a \TPR of~\priceNewTPR and a \FNR of~\priceNewFNR in the \textit{Attack} dataset, with false negatives resulting from low-profit margin. In non-attack datasets, the \acp{TNR} are~\priceNewGasTNR and~\priceNewRandTNR for \textit{High-Gas} and \textit{Regular} respectively. The false positives in these datasets mainly arise from complex transactions, arbitrage activities, and add/remove liquidity operations.}
\label{tab:price-manipulation-eval}
\rowcolors{2}{gray!10}{white}
\scalebox{0.78}{%
\begin{tabular}{l|c|cc}
\toprule
            &  & \multicolumn{2}{c}{Non-attack} \\ 
Dataset     & \textit{Attack} & \textit{High-Gas} & \textit{Regular} \\ 
\midrule
Size        & \priceAttackTxs & 1,075  & 19,989 \\ 
Gas Cost       & 6.89 $\pm$ 3.37M & 2.14 $\pm$ 1.38M & 0.24 $\pm$ 0.26M \\ 
Traversal Time & 47 $\pm$ 23ms  & 10 $\pm$ 24ms & 2.4 $\pm$ 1.5ms \\ 
TP (\%)             & \priceNewTP (\priceNewTPR)  & - & -\\
FN (\%)             & \priceNewFN (\priceNewFNR)    & - & - \\
\hspace{0.5em}\small{Low profit} & 2/2   & - & - \\
TN (\%)             & -       & \priceNewGasTN (\priceNewGasTNR) & \priceNewRandTN (\priceNewRandTNR)\\
FP (\%)             & -       & \priceNewGasFP (\priceNewGasFPR) & \priceNewRandFP (\priceNewRandFPR) \\
\hspace{0.5em}\small{Arbitrage} & -   & 0/16 & 29/103 \\
\hspace{0.5em}\small{Complex DeFi} & -   & 10/16 & 2/103 \\
\hspace{0.5em}\small{Add/Remove liquidity} & -   & 6/16 & 72/103 \\
\bottomrule
\end{tabular}
}
\end{table}

Table \ref{tab:price-manipulation-eval} presents the evaluation results in identifying price manipulation attacks. In the \textit{Attack} dataset, \framework successfully identifies~\priceNewTP malicious transactions, missing only~\priceNewFN attack transactions. This results in a \FNR of~\priceNewFNR.
In the \textit{High-Gas} and \textit{Regular} datasets, \framework achieves a \FPR of~\priceNewGasFPR and~\priceNewRandFPR, respectively.

\paragraph{Price Manipulation Refinements}
In response to observations from real-world DeFi activities and attacks, we implement two critical refinements to improve the accuracy of our traversal rule. 

Firstly, complex DeFi applications, notably those involving swaps and flash-swaps, often inadvertently trigger the price manipulation rule as outlined in Section~\ref{sec:price-oracle}.
For example,  the \texttt{doHardWork} function, interacting with Harvest Finance contracts,\footnote{An example transaction of the \texttt{doHardWork} function: 
\etherscanTx{0xcb4e7c976b4751cd93e758001135612bdd3da276b2f81814c924391d7e985f55}} executes complex actions that, despite a small swap amount, activate the price manipulation rule.
It is crucial to note that most price manipulation attacks follow a \emph{swap-and-borrow} pattern, wherein significant influence on the relative price through swaps is essential for successful exploitation.
This pattern is generally not characteristic of regular DeFi activities.
Therefore, in our refinement, we flag swap pool contracts within the transaction call traces and analyze the relative price fluctuations of these pools.
Given the relative price fluctuations, we can detect any irregular price shifts, and exclude transactions that fall in the normal range.
We accomplish this by determining the proportion of the alteration in the token balance within the swap pool relative to the aggregate token balance.

Furthermore, specific DeFi actions, such as arbitrage~\cite{zhou2021a2mm}, align with the traversal rule but operate within a reasonably controlled profit margin.
In our refinement, we calculate the absolute USD value changes of swap pools by utilizing the asset flows in \EPG and historical token prices.
By monitoring and evaluating the absolute value changes, we can effectively differentiate between legitimate transactions and potential price manipulation attempts.
This process is especially effective in false positive cases in small-scale arbitrage transactions and intricate DeFi transactions, as the compounding process only trades existing profits.

\paragraph{False Negative Analysis}

In the \textit{Attack} dataset, two of the false negatives arise from the refinement rule regarding attack profitability. Although these two transactions are initially identified by our price manipulation traversal, the profit generated is relatively low. Consequently, they are erroneously classified as non-attack transactions.

\paragraph{False Positive Analysis}
In the \textit{High-Gas} dataset,~$10$ of the false positive cases result from large-scale swap operations in complex DeFi transactions, including \textit{deposit}, \textit{withdraw} and \textit{compounding} while the other~$6$ transactions are categorized as \textit{add/remove liquidity}.

In the \textit{Regular} dataset, the false positive cases fall into two categories: \textit{add/remove liquidity} and \textit{arbitrage}, accounting for approximately~$70\%$ and~$28\%$ among the false positive cases, respectively. The former type contains actions of adding liquidity to or removing liquidity from swap pools, which performs swap operations at the beginning. The latter type encompasses large-scale \textit{arbitrage} transactions with significant earnings. These transactions profit from the price difference between swap pools. 

\subsection{RQ2: Traversal Performance Overhead}

Our time performance evaluation focuses on the overhead associated with graph traversals. As demonstrated in Tables~\ref{tab:reentrancy-eval} and~\ref{tab:price-manipulation-eval}, for attack transactions, \framework completes detection within an average time of~$270$ms. In the \textit{Regular} dataset, traversal time averages below~$100$ms, with a majority ($89.6\%$) taking less than~$20$ms.
When averaged across both attack and non-attack transactions for the two evaluated vulnerabilities, the traversal time amounts to~\averageTraversalTime per transaction.
It is crucial to mention that our prototype has not been optimized for performance.
For reference, the block interval of Ethereum is~$12$s. Our evaluation highlights the efficacy of our traversal approach, demonstrating its strong potential for real-time intrusion detection.

\subsection{RQ3: Comparison to SOTA}
\subsubsection{Comparison with SOTA Reentrancy Detection Tools}
We apply our reentrancy detection methodology on the comparison dataset (cf.\ Section~\ref{sec:benchmark-dataset}) to compare \framework with \SOTA dynamic analysis tools for reentrancy, namely \textsc{Sereum}~\cite{rodler2019sereum} and \textsc{TxSpector}~\cite{zhang2020txspector}.
The results are outlined in Table~\ref{tab:reentrancy-comparison}.
In this comparison, \textsc{Sereum}'s results are used as the baseline for ground truth.
While \textsc{Sereum} demonstrates low \acp{FNR}, \framework's FPR is substantially lower than that of \textsc{Sereum} and \textsc{TxSpector}.
Moreover, \framework surpasses both \textsc{Sereum} and \textsc{TxSpector} in terms of efficiency.
On the comparison dataset, \framework completes its detection process in an average of $30$ms per transaction, compared to \textsc{Sereum}'s $217$ms and \textsc{TxSpector}'s $1.03$s.

\begin{table}[t]
\centering
\caption{Reentrancy comparison. \framework demonstrates comparable accuracy with State-of-the-Art reentrancy detection tools \textsc{Sereum} and \textsc{TxSpector}, and superiorly outperforms both in \FPR.}\label{tab:reentrancy-comparison}
\rowcolors{2}{gray!10}{white}
\scalebox{0.78}{%
\begin{threeparttable}
\begin{tabular}{l|c|c|c}
\toprule
System     & \textsc{Sereum} \cite{rodler2019sereum} & \textsc{TxSpector \cite{zhang2020txspector}} & \textsc{Clue} \\ 
\midrule
Detection Time  & 217 $\pm$ 101ms & 1.03s (99\%{} $<$ 4s) & 30 $\pm$ 1683ms \\
\#{} Detected (TP + FP)   & 49,080          & 3,357                 & 2,323 \\
\#{} Confirmed Attacks\tnote{a} & \multicolumn{3}{c}{2,332} \\
TP\%{}\tnote{a} & 100\%{}         & 99.36\%{}             & 99.61\%{} \\
FN\%{}\tnote{a} & 0\%{}           & 0.64\%{}              & 0.39\%{} \\
TN\%{}          & 99.94\%{}       & 99.97\%{}             & 100\%{} \\
FP\%{}          & 0.06\%{}        & 0.03\%{}              & 0\%{} \\
\bottomrule
\end{tabular}
\begin{tablenotes}
  \item[a] \footnotesize{Since the ground truth is not available for such a big dataset, we assume \textsc{Sereum} does not have false negatives for comparison purposes. Among 49,080 transactions marked as positive by \textsc{Sereum}, 2,323 transactions are confirmed as real attacks.}
\end{tablenotes}
\end{threeparttable}
}
\end{table}
\begin{table}[t]
\centering
\caption{Price manipulation comparison. \framework superiorly outperforms State-of-the-Art price manipulation detection tools in \TPR.}\label{tab:price-manipulation-comparison}

\rowcolors{2}{gray!10}{white}
\scalebox{0.78}{%
\begin{threeparttable}
\begin{tabular}{l|c|c|c|c|c}
\toprule
System     & \textsc{DeFiRanger}~\cite{wu2023defiranger} & \textsc{FlashSyn}~\cite{chen2024flashsyn} & \textsc{DeFiTainter}~\cite{kong2023defitainter} & \textsc{DeFort}~\cite{xie2024defort} & \textsc{Clue} \\ 
\midrule
\#{} Attacks  & \multicolumn{5}{c}{54} \\
\#{} Detected  & 4          & 8 & 27 & 52 & \textbf{54} \\
TP\%{} & 7.41\%{} & 14.81\%{} & 50.00\%{} & 96.30\%{} & \textbf{100.00\%{}} \\
FN\%{} & 92.59\%{} & 85.19\%{} & 50.00\%{} & 3.70\%{} & \textbf{0.00\%{}} \\
\bottomrule
\end{tabular}
\end{threeparttable}
}


\end{table}

\subsubsection{Comparison with SOTA Price Manipulation Detection Tools}

We evaluated the performance of our price manipulation detection methodology against four \SOTA price manipulation detection systems, namely \textsc{DeFiRanger}~\cite{wu2023defiranger}, \textsc{FlashSyn}~\cite{chen2024flashsyn}, \textsc{DeFiTainter}~\cite{kong2023defitainter}, and \textsc{DeFort}~\cite{xie2024defort}.
We used the $D1$ cenchmark dataset proposed by \textsc{DeFort}~\cite{xie2024defort} as mentioned in Section~\ref{sec:benchmark-dataset}.
As shown in Table~\ref{tab:price-manipulation-comparison}, \framework achieves superior accuracy, identifying all 54 attacks with a true positive rate of 100\%{} and no false negatives.
In contrast, DeFort~\cite{xie2024defort}, the second-best performer, achieves a 96.30\%{} true positive rate while missing 3.70\%{} of attacks.
Unfortunately, a complete comparison on the \FPR and detection speed is not feasible, as these tools either lack open-source implementations or do not report these metrics in their evaluations.
While not directly comparable due to different evaluation datasets, \textsc{DeFort}~\cite{xie2024defort} reports a detection time of $10 \sim 200$ms per transaction, whereas \framework{} achieves an average detection time of $2.9 \pm 5.72$ms on our dataset, suggesting promising efficiency in price manipulation detection.

\revision{\subsection{RQ4: Ablation Study}}

\revision{
To validate the necessity of \EPG's unified structure, we conduct an ablation study in which each constituent graph is omitted. Table~\ref{tab:ablation} presents the results: removing certain graphs degrades efficiency or accuracy, while others render the traversals non-executable. These outcomes reinforce the importance of integrating all three graphs into a single \EPG for reliable analysis.

\subsubsection{Omitting \CTG}
Because \CTG serves as the backbone of \EPG, connecting all contract interactions, completely removing it from \EPG is not feasible. Therefore, we conduct our ablation study by omitting only the asset flow information from \CTG, while retaining the contract vertices and \texttt{CALL} edges (cf.\ Figure~\ref{fig:epg}).

\paragraph{Reentrancy} Omitting asset flow information forces the reentrancy traversal to consider all data sources, not just those related to asset flows in \CTG. While this does not affect \framework's accuracy on our evaluation dataset, the average detection time on the \textit{High-Gas} dataset increases from~$52\pm109$ms to~$179\pm\numprint{2708}$ms. This result demonstrates that \CTG's asset flow information is crucial for analysis efficiency, as its removal leads to examining irrelevant data dependencies and significant computational overhead.

\paragraph{Price manipulation} For price manipulation detection, the results show a surge in \FPR~---~from \priceNewGasFPR to \priceAblationGasFPR on the \textit{High-Gas} dataset and from \priceNewRandFPR to \priceAblationRandFPR on the \textit{Regular} dataset. The \FNR decreases because the traversal over-flags transactions as attacks, capturing all true positives but at the cost of massive false positives. This \FPR spike is due to the loss of causal links between storage manipulations (via \DDG) and financial outcomes (via \CTG), as the analysis cannot distinguish storage updates that cause abnormal price changes from normal operations. These results confirm that \CTG is essential for modeling inter-contract interactions and asset flow dynamics.

\subsubsection{Omitting \DDG} In the experiments omitting \DDG, we remove the storage vertices along with the \texttt{CONTROL} and \texttt{WRITE} edges from the \EPG.

\paragraph{Reentrancy} To run our reentrancy traversal on the graph with the \DDG omitted, we modify the traversal by removing the \textsc{ControlBlock} rule (cf.\ Section~\ref{sec:reentrancy}). The results show an increased \FPR on both the \textit{High-Gas} dataset (from \reentrancyNewGasFPR to \reentrancyAblationGasFPR) and the \textit{Regular} dataset (from \reentrancyNewRandFPR to \reentrancyAblationRandFPR). The false positives primarily stem from benign transactions involving callback functions (e.g., flash loans), where the \CTG and \DCFG alone prove insufficient to distinguish between legitimate nested calls and malicious reentrancy attacks. The \DDG plays a critical role in verifying whether reentrant calls are controlled by stale storage states via its data dependencies.
\paragraph{Price manipulation} As discussed in Section~\ref{sec:price-oracle}, the \textsc{WriteControl} traversal rule for price manipulation primarily relies on identifying storage changes (provided by the \DDG) that ``influence'' asset transfers. Therefore, omitting the \DDG renders \textsc{WriteControl} non-executable.

\subsubsection{Omitting \DCFG} As shown in Figure~\ref{fig:epg}, the basic block vertices from the \DCFG connect contract vertices from the \CTG with storage vertices from the \DDG. Without the \DCFG, this graph cannot be formed. Moreover, omitting the \DCFG invalidates the \textsc{ControlBlock} and \textsc{SuccBlock} rules for reentrancy traversal (cf. Figure~\ref{fig:reentrancy_traversal}) as well as the \textsc{WriteControl} rule for price manipulation (cf. Figure~\ref{fig:price_traversal}). Consequently, both traversals become non-executable when the \DDG is omitted.
}

\begin{table}[t]
\centering
\caption{\revision{Ablation Study Results. Omitting asset flow information from the \CTG leads to an increased average detection time for reentrancy and a rise in the \FPR for price manipulation. Omitting the \DDG degrades the accuracy of reentrancy detection and renders price manipulation traversal non-executable. Finally, the removal of the \DCFG results in both traversals becoming non-executable.}}
\label{tab:ablation}
\rowcolors{2}{gray!10}{white}
\resizebox{\columnwidth}{!}{%
\begin{threeparttable}
\begin{tabular}{lccccccc}
\toprule
\multirow{2}{*}{\textbf{Ablation}} 
  & \multicolumn{4}{c}{\textbf{Reentrancy}} 
  & \multicolumn{3}{c}{\textbf{Price Manipulation}} \\
\cmidrule(lr){2-5}\cmidrule(lr){6-8}
 & \makecell{Attack\\(TP\%/FN\%)} 
 & \makecell{High-Gas\\(TN\%/FP\%)} 
 & \makecell{Regular\\(TN\%/FP\%)} 
 & \makecell{Detection Time\tnote{a}\\(ms)} 
 & \makecell{Attack\\(TP\%/FN\%)} 
 & \makecell{High-Gas\\(TN\%/FP\%)} 
 & \makecell{Regular\\(TN\%/FP\%)} \\
\midrule
Full \EPG (Baseline)      
 & \reentrancyNewTPR/\reentrancyNewFNR 
 & \reentrancyNewGasTNR/\reentrancyNewGasFPR
 & \reentrancyNewRandTNR/\reentrancyNewRandFPR
 &  $52 \pm 109$
 & \priceNewTPR/\priceNewFNR 
 & \priceNewGasTNR/\priceNewGasFPR
 & \priceNewRandTNR/\priceNewRandFPR \\
Omit \CTG (Asset Flow)   
 & \reentrancyNewTPR/\reentrancyNewFNR
 & \reentrancyNewGasTNR/\reentrancyNewGasFPR
 & \reentrancyNewRandTNR/\reentrancyNewRandFPR
 &  $179 \pm \numprint{2708}$
 & \priceAblationAttackTPR/\priceAblationAttackFNR
 & \priceAblationGasTNR/\priceAblationGasFPR
 & \priceAblationRandTNR/\priceAblationRandFPR \\
Omit \DDG   
 & \reentrancyNewTPR/\reentrancyNewFNR 
 & \reentrancyAblationGasTNR/\reentrancyAblationGasFPR
 & \reentrancyAblationRandTNR/\reentrancyAblationRandFPR
 &  $48 \pm 81$
 & \multicolumn{3}{c}{Traversal Not Executable} \\
 Omit \DCFG   
 & \multicolumn{4}{c}{Traversal Not Executable} & \multicolumn{3}{c}{Traversal Not Executable}\\
\bottomrule
\end{tabular}%
\begin{tablenotes}
\item[a] \small{Average detection time on the \textit{High-Gas} dataset. We report the detection time on the \textit{High-Gas} dataset as it better represents transactions with complex execution logic (cf.\ Table~\ref{tab:evaluation-dataset}); omitting \CFG results in a slight increase in detection time on the other datasets as well.}
\end{tablenotes}
\end{threeparttable}
}
\end{table}

\revision{
\section{Generalizability}\label{sec:generalizability}
The \EPG unifies three fundamental perspectives on smart contract execution: \emph{call traces} (\CTG) for inter-contract invocation and asset flows, \emph{dynamic control flow} (\DCFG) for branching and function-level execution paths, and \emph{data dependency} (\DDG) for state updates and variable interactions. By merging these views, \EPG vertices and edges inherently encode comprehensive details of each transaction's execution state. This integrated design allows \framework to detect a wide range of vulnerabilities that depend on call order, storage variables, or asset transfer patterns, including reentrancy and price manipulation.
 
Because \EPG encodes both control- and data-flow information alongside asset transfers, \framework can diagnose vulnerabilities where state changes (in \DDG) are coupled with execution paths (\DCFG) or cross-contract interactions (\CTG). Classic reentrancy attacks rely on dynamic checks across multiple calls, and price manipulation exploits subtle token-value updates. \EPG's structure accommodates these and other commonly exploited patterns in \DeFi. Whenever an attack's root cause involves data or control dependencies and asset movements~---~such as a storage variable influencing trade outcomes~---~\EPG traversals can likely capture it by design. We emphasize that \framework's ``generalizability'' does not imply any single traversal rule automatically detects every vulnerability. Rather, it offers a unified approach to capturing the rich details of smart contract executions (via the \EPG) and efficiently analyzing them (through graph traversal), enabling security experts to swiftly address emerging threats.
 
Despite its breadth, some attacks demand additional domain-specific information or specialized checking logic. For instance, our price manipulation detection (cf.\ Section~\ref{sec:price-manipulation-rq1}) integrates swap pool information. Embedding these additional data in \EPG extends \framework's coverage with minimal engineering overhead, thanks to \EPG's unified representation and extensibility. In practice, security experts simply augment vertex properties or refine edge relationships and then update the corresponding traversal logic.

While incorporating new attacks in \framework requires understanding their intrinsic patterns and formalizing them into traversal rules~---~an effort that inevitably relies on expert insight~---~\framework  streamlines this process through its unified data representation. By centralizing rich execution details within the \EPG, experts can concentrate on specifying the exploit's conditions without re-implementing instrumentation or reconstructing partial graphs. This significantly accelerates the analysis, especially when newly disclosed vulnerabilities require urgent investigation to avert further exploits. This approach has already enabled \EPG to handle read-only reentrancy with only small refinements to its reentrancy traversal as detailed Section~\ref{sec:case-study-read-only}.

\subsection{Case Study: Read-only Reentrancy}\label{sec:case-study-read-only}
In the following, we illustrate \framework's generalizability by extending it to detect a novel reentrancy variant known as \emph{read-only reentrancy}, which has eluded \SOTA tools.
Unlike conventional reentrancy, which involves state modifications, read-only reentrancy exploits the potential for view functions to return outdated information during reentrant calls.

\subsubsection{Traversal} \framework identifies read-only reentrancy by extending the \textsc{ControlBlock} traversal (cf. Section~\ref{sec:reentrancy}) as follows. It considers not only write operations but also read operations affected by stale data sources. Additionally, since the victim asset flow may occur outside the reentrant view function call, the traversal examines \texttt{CONTROL} edges beyond the reentrant call. This extension does not require any changes to the underlying \EPG, but only minimal adjustments to the traversal rules on line~\ref{line:controlblock_start} and~\ref{line:controlblock_end} of Algorithm~\ref{alg:reentrancy}.

In contrast, \SOTA tools like \textsc{Sereum}~\cite{rodler2019sereum} and \textsc{TxSpector}~\cite{zhang2020txspector} would require foundational modifications to detect read-only reentrancy.
Specifically, \textsc{Sereum}~\cite{rodler2019sereum} would need to redesign its taint patterns and state access tracking mechanisms, while \textsc{TxSpector}~\cite{zhang2020txspector} would have to introduce new predicates and detection rules in its logic system.
\framework's adaptable traversal-based approach, however, allows for seamless extension to capture this variant.

\begin{table}[t]
\centering
\caption{Read-only reentrancy evaluation.}\label{tab:reentrancy-read-eval}

\begin{subtable}[t]{0.39\textwidth}
    \centering
    \caption{Read-only reentrancy comparison. \framework is currently the only dynamic analysis tool that can detect read-only reentrancy.}
    \scalebox{0.8}{%
    \rowcolors{2}{gray!10}{white}
    \begin{tabular}{l|cc}
    \toprule
    Victim Protocol & sentiment.xyz & dForce \\
    \midrule
    \textsc{Sereum} \cite{rodler2019sereum} & \xmark & \xmark \\
    \textsc{TxSpector} \cite{zhang2020txspector} & \xmark & \xmark \\
    \framework & \cmark & \cmark \\
    \bottomrule
    \end{tabular}}
\end{subtable}%
\hfill
\begin{subtable}[t]{0.58\textwidth}
    \centering
    \caption{\framework achieves a \FNR of \reentrancyReadNewFNR on the two documented read-only reentrancy attacks. In non-attack datasets, the \TNR is \reentrancyReadNewGasTNR and \reentrancyReadNewRandTNR for \textit{High-Gas} and \textit{Regular} respectively.}
    \scalebox{0.8}{%
    \rowcolors{2}{gray!10}{white}
    \begin{tabular}{l|c|cc}
    \toprule
                &  & \multicolumn{2}{c}{Non-attack} \\ 
    Dataset     & \textit{Attack} & \textit{High-Gas} & \textit{Regular} \\ 
    \midrule
    Size            & 2 & \reentrancyGasTxs  & \reentrancyRandTxs \\ 
    Gas Cost       & 5.97 $\pm$ 1.84M & 2.13 $\pm$ 1.38M & 0.24 $\pm$ 0.29M \\
    Traversal Time & 4.02 $\pm$ 0.58s  & 48 $\pm$ 768ms & 12 $\pm$ 70ms \\ 
    TN (\%)             & -       & \reentrancyReadNewGasTN (\reentrancyReadNewGasTNR) & \reentrancyReadNewRandTN (\reentrancyReadNewRandTNR)\\
    FP (\%)             & -       & \reentrancyReadNewGasFP (\reentrancyReadNewGasFPR) & \reentrancyReadNewRandFP (\reentrancyReadNewRandFPR)\\
    \bottomrule
    \end{tabular}}
\end{subtable}
\end{table}

\subsubsection{Evaluation}
To evaluate our read-only reentrancy traversal, we expand the evaluation dataset (cf.\ Section~\ref{sec:evaluation-dataset}) to include two documented attack transactions from the sentiment.xyz~\cite{mediumDecodingSentiment} and dForce~\cite{dforceAttack} incidents.\footnote{Note that read-only reentrancy is a recently identified vulnerability, and consequently, there are few documented incidents.} Table~\ref{tab:reentrancy-read-eval} presents the evaluation results. \framework achieves a~$100\%$ \TPR on the \textit{Attack} dataset, detecting both known incidents. The \FPR remains low, at~$0.37\%$ ($4$ out of~$\numprint{1077}$) for the \textit{High-Gas} dataset and~$0.28\%$ ($55$ out of~$\numprint{19985}$) for the \textit{Regular} dataset.

The evaluation results highlight that the read-only reentrancy detection with \framework is both simple and effective. This case study demonstrates that with minimal engineering, \framework rapidly adapts to emerging threats, outperforming \SOTA solutions like \textsc{Sereum}\cite{rodler2019sereum} and \textsc{TxSpector}\cite{zhang2020txspector}, which fail to identify these attacks.

}

\section{Related Work}
\subsection{Graph-based Static Program Analysis}
Graphs have become the fundamental building blocks in the field of program analysis. In tasks such as program optimization and vulnerability discovery, the utilization of various types of graphs, including \AST, \CFG, and \PDG, is essential for achieving accurate and reliable results. Combining \AST, \CFG, and \PDG, Yamaguchi \etal~\cite{yamaguchi2014modeling} first introduce the concept of \CPG, which represents program source code as a property graph. Such a comprehensive view of code enables rigorous identification of vulnerabilities through graph traversals. \CPG has been shown to be effective in identifying buffer overflows, integer overflows, format string vulnerabilities, and memory disclosures~\cite{yamaguchi2014modeling}. Giesen \etal~\cite{giesen2022practical} apply the \CPG approach to smart contracts and propose hardening contract
compiler (HCC). HCC models control-flows and data-flows of a given smart contract statically as a \CPG, which allows efficient detection and mitigation of integer overflow and reentrancy vulnerabilities.
Pasqua \etal~\cite{pasqua2023enhancing} and Contro \etal~\cite{contro2021ethersolve} investigate the generation of precise and accurate static \CFG from \EVM bytecode using symbolic execution. Such approaches significantly enhance the accuracy of security analyses.
Inspired by the previous studies, we propose the \EPG to model the dynamic contract execution details as a property graph. Compared to the static approaches, the \EPG captures runtime information exposed from the concrete executions and hence complements the contract security analysis, particularly in the online and postmortem scenarios.

\subsection{Smart Contract Dynamic Analysis}
Previous studies in the field of smart contract dynamic analysis have primarily focused on two key directions, \textit{(i)} online attack detection and \textit{(ii)} forensic analysis.

\subsubsection{Online Attack Detection} Grossman \etal~\cite{grossman2017online} develop a polynomial online algorithm for checking if an execution is effectively callback free, a property for detecting reentrancy vulnerabilities. Rodler \etal~\cite{rodler2019sereum} introduce \textsc{Sereum}, a runtime solution to detect smart contract reentrancy vulnerabilities. \textsc{Sereum} exploits dynamic taint tracking to monitor data-flows during contract execution and applies to various types of reentrancy vulnerabilities. Chen \etal~\cite{chen2020soda} develop \textsc{SODA}, an online framework for detecting various smart contract attacks. Torres \etal~\cite{ferreira2020aegis} propose a dynamic analysis tool for the \EVM based on a domain-specific language. To the best of our knowledge, \framework represents the first generic dynamic analysis framework for the EVM that offers precise data flow tracking capabilities.

\subsubsection{Forensic analysis} Perez and Livshits~\cite{perez2021smart} propose a Datalog-based formulation for performing analysis over \EVM execution traces and conduct a large-scale evaluation on~$\numprint{23327}$ vulnerable contracts. Zhou \etal~\cite{zhou2020ever} undertake a measurement study on $420$M Ethereum transactions, constructing transaction trace into action and result trees. The action tree gives information of contract invocations, while the result tree provides asset transfer data, which are then compared against predefined attack patterns. Zhang \etal~\cite{zhang2020txspector} design~\textsc{TxSpector}, a logic-driven framework to investigate Ethereum transactions for attack detection. \textsc{TxSpector} encodes the transaction trace into logic relations and identifies attacks following user-specified detection rules. The primary distinction between \textsc{TxSpector} and \framework lies in their respective approaches: \textsc{TxSpector} employs logic programming, whereas \framework utilizes a graph-based method. As indicated in Table~\ref{tab:reentrancy-comparison}, our findings suggest that graph traversals offer greater efficiency compared to logic programming techniques. Furthermore, the \EPG can easily enrich existing transaction explorers (e.g., \url{https://openchain.xyz/trace}) with more detailed dynamic execution information, which may significantly facilitate manual forensic analysis. Tools for detecting price manipulation attacks have been studied in~\cite{wu2023defiranger,chen2024flashsyn,kong2023defitainter,xie2024defort}. Our evaluation demonstrates that \framework outperforms these tools in terms of \TPR. \revision{Eshghie \etal~\cite{eshghie2021dynamic} propose Dynamit to extract features from transaction data and use machine learning models to detect reentrancy attacks. Given that \EPG captures rich transaction execution details, it has potential to serve as valuable features for machine learning approaches, which we leave for future work.}

\subsection{Smart Contract and \DeFi Attacks}
There has been a growing body of literature examining the prevalence of smart contract attacks, with a particular emphasis on those targeting DeFi platforms. Qin~\etal~\cite{qin2021attacking} study the first two \DeFi attacks and propose a numerical optimization framework that allows optimizing attack parameters. Li \etal~\cite{li2022survey} conduct a comprehensive analysis of the real-world \DeFi vulnerabilities. Zhou \etal~\cite{zhou2022sok} present a reference framework that categorizes $181$ \DeFi attacks occurring between~$2018$ and~$2022$, related academic papers, as well as security audit reports into a taxonomy. Zhang \etal~\cite{zhang2023demystifying} analyze~$516$ real-world smart contract vulnerabilities from~$2021$ to~$2022$, categorize undetectable bugs and offer insights into their causes, effects, and mitigation strategies. 
The related work~\cite{zhou2022sok,zhang2023demystifying} highlights that some attacks are executed over multiple transactions, a method attackers might use to evade \SOTA intrusion prevention systems~\cite{qin2023blockchain}. It is crucial to note that \framework, at present, lacks the capability to detect attacks spanning multiple transactions. Enhancing \framework to address this limitation is earmarked for future development.

\section{Conclusion}
In this paper, we introduce a generic dynamic analysis framework \framework designed for the \EVM. \framework employs a novel approach centered around the \EPG and an innovative graph traversal technique. Together, these elements provide an efficient and effective method for identifying potential smart contract attacks. Our evaluations demonstrate \framework's exemplary performance, showcasing high \acp{TPR} and low \acp{FPR} in detecting reentrancy and price manipulation, thus outperforming existing dynamic analysis tools. The efficiency of \framework renders it a valuable tool for conducting comprehensive forensic analysis as well as facilitating real-time intrusion detection. 
This work represents a significant advancement in enhancing smart contract security and contributes valuable new tools for transaction security analysis in the complex landscape of \DeFi.

\section*{Data Availability}

Our artifact and dataset are available at \url{https://github.com/sunblaze-ucb/clue}.

\begin{acks}
This material is in part based upon work supported by the Center for Responsible, Decentralized Intelligence at Berkeley (Berkeley RDI) and partially supported by an Ethereum Foundation Academic Grant. Any opinions, findings, and conclusions or recommendations expressed in this material are those of the author(s) and do not necessarily reflect the views of these institutes.
\end{acks}

\bibliographystyle{ACM-Reference-Format}
\bibliography{references}


\end{document}